\documentclass[useAMS,usenatbib]{mn2e}
\usepackage{graphicx}
\usepackage{multirow}

\title[The $\beta$ Cep/SPB star 12 Lacertae]{The $\beta$ Cep/SPB star 12 Lacertae: extended mode identification and complex seismic modelling}
\author[Daszy\'nska-Daszkiewicz, Szewczuk \& Walczak]{J. Daszy\'nska-Daszkiewicz$^{1}$\thanks{E-mail:
daszynska@astro.uni.wroc.pl}, W. Szewczuk$^{1}$\thanks{E-mail:
szewczuk@astro.uni.wroc.pl} and P. Walczak$^{1}$\thanks{E-mail:
walczak@astro.uni.wroc.pl}\\
$^{1}$Instytut Astronomiczny Uniwersytet Wroc{\l}awski, Wroc{\l}aw, Poland}

\begin{document}

\date{}

\pagerange{\pageref{firstpage}--\pageref{lastpage}} \pubyear{}

\maketitle

\label{firstpage}

\begin{abstract}
Results of mode identification and seismic modelling of the $\beta$ Cep/SBP star 12 Lacertae are presented.
Using data on the multi-colour photometry and radial velocity variations, we determine or constrain
the mode degree, $\ell$, for all pulsational frequencies. Including the effects of rotation, we show that
the dominant frequency, $\nu_1$, is most likely a pure $\ell=1$ mode and the low frequency, $\nu_A$,
is a dipole retrograde mode. We construct a set of seismic models which fit two pulsational
frequencies corresponding to the modes $\ell= 0,$ p$_1$ and $\ell= 1,$ g$_1$ and reproduce also the complex amplitude
of the bolometric flux variations, $f$, for both frequencies simultaneously.
Some of these seismic models reproduce also the frequency $\nu_A$, as a mode $\ell= 1,$ g$_{13}$ or g$_{14}$,
and its empirical values of $f$. Moreover, it was possible to find a model fitting the six 12 Lac frequencies (the first five and $\nu_A$),
only if the rotational splitting was calculated for a velocity of $V_{\rm rot}\approx 75$ km/s.
In the next step, we check the effects of model atmospheres, opacity data, chemical mixture and opacity enhancement. Our results show that the OP tables are preferred
and an increase of opacities in the $Z-$bump spoils the concordance of the empirical and theoretical values of $f$.
\end{abstract}
\begin{keywords}
stars: early-type -- stars: oscillations -- stars: rotation -- stars: individual: 12 Lac -- atomic data: opacities
\end{keywords}

\section{Introduction}
The puzzle about the origin of pulsation instability in the B-type main sequence stars had persisted for many years.
The explanation of this phenomenon is only 20 years old (Cox et al. 1992, Kiriakidis et al. 1992, Moskalik \& Dziembowski 1992,
Dziembowski \& Pamyatnykh 1993, Gautschy \& Saio 1993) and has been possible after re-computation of the opacity data,
which included a huge number of transitions in the heavy element ions (Iglesias et al. 1992, Seaton 1993).
These new opacities revealed the metal opacity bump
around $2\cdot 10^5$ K, which is called the $Z-$ or Fe$-$bump, because transitions within the iron M-shell dominate.
Since then, many intriguing results have been obtained for the B-type pulsators,
mainly thanks to data from the multi-site photometric and spectroscopic campaigns organized
for several of such objects as well as from the space-based missions like MOST \citep{MOST}, CoRoT \citep{Corot} or Kepler \citep{Kepler}.
One of the most important results was the discovery of high-order g-modes in early B-type pulsators
in which typically low order p/g modes are observed. These variables were termed the $\beta$ Cep/SPB stars or hybrid pulsators
and a few of them were detected up to now, e.g., $\nu$ Eridani (Handler et al. 2004, Aerts et al. 2004, Jerzykiewicz et al. 2005),
12 Lacertae (Handler et al. 2006; Desmet et al. 2009) or $\gamma$ Pegasi (Handler 2009, Handler et al. 2009).
Since the release of the first revised opacity tables many improvements have been added
in computation of these atomic data (Iglesias \& Rogers 1996, Seaton 1996, 2005). Moreover, the solar chemical mixture has been redetermined several times (Grevesse \& Noels 1993,	Grevesse \& Sauval 1998, Asplund et al. 2005, 2009). Despite of that, still a common problem is to get the high-order g-modes excited in models with masses larger than about 7 $M_{\odot}$.

The hybrid pulsators give challenges as well as possibilities. Excitation of both low order p/g modes
and high order g-modes allow potentially to scan almost the whole interior of a star.
On the other hand, the problem with instability of high-order g-modes indicates that
there is still something we do not understand or cannot adequately take into account in stellar physics.
Usually, the opacity data are ,,blamed'' for the inconsistencies and an artificial increase of the opacity coefficient
was called, e.g., for the $\nu$ Eri star (Pamyatnykh, Handler \& Dziembowski 2004).

In this paper we reanalyse the frequency spectrum of the hybrid B-type pulsator 12 Lacertae. In Section\,2, we give an overview of the star.
Section\,3 is devoted to mode identification, which we made both within the zero-rotation approximation
and including some effects of rotation for frequencies which can be mostly affected by it.
In Section\,4, we present results of seismic modelling which we made in two steps, i.e., 1) fitting frequencies
and 2) fitting the nonadiabatic $f-$parameter. We study the effects of model atmospheres, opacity data,
chemical mixture and opacity enhancement. Conclusions are summarised in Section\,5.

\section{The variable star: 12 Lacertae}

12 Lacertae (DD Lac, HR 8640, HD 214993) is a well-known pulsating star of B2III spectral type.
Studies of the variability of the star have begun in the early twentieth century when
\citet{adams} discovered its radial velocity variations. The period of these changes amounted to ${\rm P}=0^{\rm d}.193089$
as determined by \citet{young}.

In the following years 12 Lac had been extensively studied by many authors, e.\,g.\,\citet{young1919}, \citet{christie}.
However, the physical interpretation of the radial velocity changes was incorrect; they thought that 12 Lac is a binary system.
\citet{stebbins} and \citet{guthnick} reported the light variations with periods of $0^d.193$ and $0^d.1936$, respectively, which agreed well with the spectroscopic value of \citet{young}.
\citet{fath1938} explained a shape of the light curve as an interference between two close periods of the order of $0^d.193$. In his further studies, \citet{fath1947} found three periods; the well-known primary $0^d.19308902$, the secondary $0^d.164850$ and the third $0^d.1316$.
However, these periods have not been confirmed by \citet{dejager1953}, who determined the new value of the primary period $0^d.1930883$
and found the secondary $0^d.197367$.
In 1956, one of the first worldwide observing campaigns was organized called The International Lacertae weeks \citep{dejager}.
The analysis of these observations by \citet{barning} led to a discovery of four different periods; two already known:
$0^d.19308883$, $0^d.197358$, and two new ones: $0^d.182127$, $25^d.85$.
\citet{jerzykiewicz1978} chose a more homogeneous sample of yellow passband data from this campaign and detected six periods: P$_1=0^d.1930754$, P$_2=0^d.19737314$, P$_3=0^d.18214713$, P$_4=0^d.1874527$, P$_5=0^d.0951112$ and P$_6=0^d.23582180$.


Sato (1979) examined observational data of 12 Lac to test stability of pulsational periods.
Next, secular variations of the period were searched by Ciurla (1987), but no variations larger than 0.08 s per century were found.
After the great effort made during The International Lacertae weeks more photometric data were gathered and analysed by
Jarzebowski et al. (1979) and Jerzykiewicz et al. (1984).
Smith (1980) examined SiIII $\lambda$ 4567 line profile variations. His findings
were generally in agreement with previous results of Jerzykiewicz (1978).
The spectroscopic analysis based on high spectral and time resolution observations
had been carried out by Mathias et al. (1994). Their frequency analysis
revealed a presence of 6 photometric frequencies in the radial velocity curve.
Subsequently \citet{dziembowski_jerzykiewicz1999} investigated the possibility of existence of the equidistant
triplet, $\nu_1$, $\nu_4$, $\nu_3$ (corresponding to periods P$_1$, P$_4$ and P$_3$) and performed seismic modelling.

In the years 2003-2004, the two multisite campaigns for 12 Lac were organized. 
\citet{handler2006} reported results on the photometric campaign; they confirmed five from six already known frequencies
and added six new independent modes, including the low frequency typical for Slowly Pulsating B-type stars (SPB).
The frequency values determined from these data are: $\nu_1$=5.179034, $\nu_2$=5.066346, $\nu_3$=5.490167, $\nu_4$=5.334357,
$\nu_5$=4.24062, $\nu_6$=7.40705, $\nu_7$=5.30912, $\nu_8$=5.2162, $\nu_9$=6.7023, $\nu_{10}$=5.8341
and $\nu_{\rm A}$=0.35529 c/d. Based on the colour photometry \citet{handler2006} provided the spherical degrees $\ell$
for the five strongest frequency peaks and constraints on $\ell$ for the low-amplitude frequencies.
Then, \citet{desmet2009} presented results on the spectroscopic campaign. They confirmed 10 independent frequencies discovered
 from photometry. The important result was that in their spectroscopy the SPB-type frequency has been also detected.
Thus, 12 Lac is another known hybrid pulsator in which  the $\beta$ Cep and  SPB type modes are excited simultaneously.
From spectroscopy, \citet{desmet2009} arrived at a unique identification of the mode degrees, $\ell$,
and azimuthal orders, $m$, for four pulsational frequencies.

Dziembowski \& Pamyatnykh (2008), based on results from the 2003 world-wide campaign, made seismic modelling of 12 Lac
and showed that  the rotation rate increases significantly towards the star centre. They addressed also a problem with
excitation of low frequency mode $\nu_A$.

\begin{figure}
\includegraphics[width=\columnwidth]{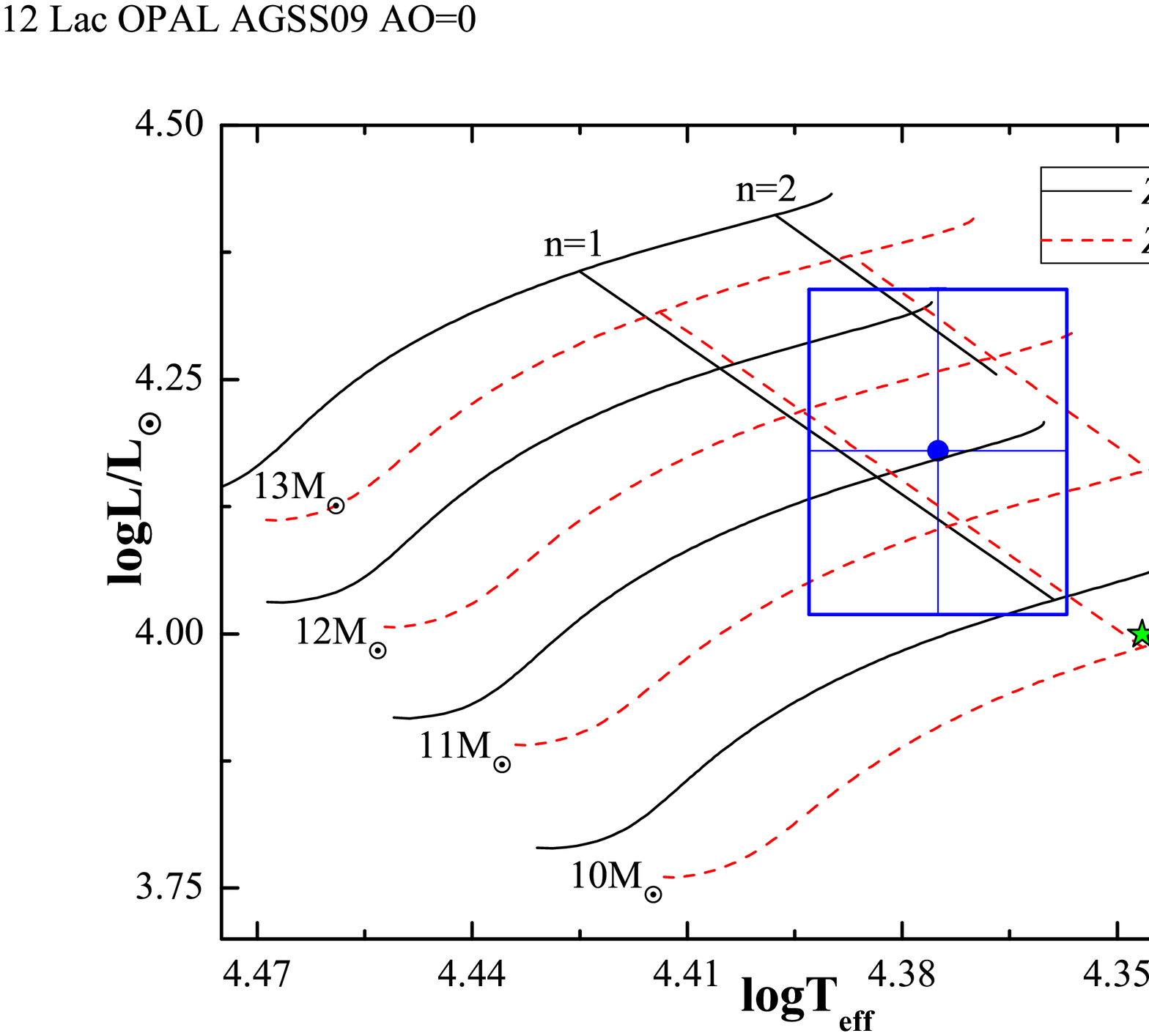}
 \caption{The HR diagram with a position of the observational error box of 12 Lac as determined by \citet{handler2006}.
 The evolutionary tracks were computed for two values of metallicity, $Z=$0.010, 0.015, the hydrogen abundance of $X=0.7$,
 initial rotational velocity of $ V_{\rm rot}=50$ km/s, no overshooting from the convective core and the OPAL tables.
 Lines labeled as $n=1$, $n=2$ and an asterisk will be discussed latter in the text.}
 \label{hr}
\end{figure}

In Fig.\,\ref{hr}, we show the observational error box of 12 Lac in the HR diagram. The star is in an advanced phase
of the core hydrogen burning.
We adopted the effective temperature, $\log T_{\rm eff}=4.375 \pm0.018$ and luminosity, $\log L/L_\odot=4.18 \pm 0.16$
from \citet{handler2006}.
In the literature various determinations of metallicity are given. Niemczura \& Daszy\'nska-Daszkiewicz (2005) obtained
[m/H]$\approx -0.2\pm 0.1$ from the ultraviolet IUE spectra, which corresponds to $Z\approx 0.012\pm 0.003$,
whereas \citet{morel2006} derived $Z=0.0089 \pm 0.0018$ from the high resolution optical spectra.
Therefore, we show evolutionary tracks for two values of metallicity:  $Z=0.015$ and  $Z=0.010$.
Computations were performed using the Warsaw-New Jersey evolutionary code (Pamyatnykh et al. 1998), with the OPAL equation
of state \citep{RN02}, the OPAL opacity tables \citep{opal} and the latest heavy element mixture by \citet{asplund2009},
hereafter AGSS09. For lower temperatures, the opacity data were supplemented with the Ferguson tables \citep{Fal05,Sal09}.
We assumed the hydrogen abundance of $X=0.7$, initial rotational velocity of $V_{\rm rot}=50$ km/s and no overshooting from
the convective core. Rotational velocity was chosen of the same magnitude as in \citet{desmet2009}
who derived $V_{\rm rot}\sin i=36\pm 2$ km/s and $V_{\rm rot}=49\pm 3$ km/s.
It means that 12 Lac rotates at a speed of about 10\% of its break-up velocity ($V_{\rm rot}^{\rm crit}\approx\sqrt{\frac{GM}{R}}\approx 480$ km/s).


\section[]{Mode identification}

To identify pulsational modes detected in 12 Lac, we will use the most popular observables, i.e., the amplitudes
and phases of the photometric and radial velocity variations. Here, we rely on the Str\"omgren $uvy$ photometry
\citep{handler2006} and the radial velocity changes \citep{desmet2009} determined as the first moment of
the SiIII 4552.6\AA ~line. We performed mode identification in two steps. Firstly, we determined the degree $\ell$
ignoring all effects of rotation on pulsation. This is a commonly used approach and in the case of 12 Lac, which rotates
at a speed of about 50 km/s, seems to be justified for the most frequencies.
However, this value of rotation can already couple pulsational modes and affects properties of slow modes such as
the frequency $\nu_{\rm A}$ detected in 12 Lac. Therefore, in the second step, we checked whether the dominant
frequency, $\nu_1$, can be a rotational coupled mode and apply traditional approximation to identify the angular numbers
of the low frequency, $\nu_{\rm A}$.

\subsection[]{The zero-rotation approximation}

If all effects of rotation on stellar pulsation are neglected then identification of the mode degree, $\ell$,
from photometric amplitudes and phases are independent of the inclination angle, $i$, intrinsic mode amplitude, $\varepsilon$,
and azimuthal order, $m$. This is also a disadvantage of disregarding rotation because the full geometry of mode cannot be determined.
Within the zero-rotation approximation, the complex amplitude of the light variation in the $x$ passband
in the framework of linear theory can be written as (e.g., Daszy\'nska-Daszkiewicz et al. 2002)
\begin{equation}
  \label{amplitudy}
  \mathcal{A}^x(i)=-1.086 \varepsilon Y_{\ell}^m(i,0)b_{\ell}^x(D_{1,\ell}^x+D_{2,\ell}+D_{3,\ell}^x),
\end{equation}
where $Y_{\ell}^m$ is the spherical harmonic, $b_{\ell}^x$ is disc averaging factor, and
$D_{1,\ell}^x$, $D_{2,\ell}$ and $D_{3,\ell}^x$ are the temperature, geometrical and pressure effects, respectively.
Expressions for these quantities can be found in Daszy\'nska-Daszkiewicz et al. (2002).

The radial velocity variations due to pulsations are defined as the first moment of the well isolated spectral line
and the formulae is given in Dziembowski (1977).

To compute the theoretical values of the photometric amplitudes and phases two inputs are needed.
The first one comes from models of stellar atmospheres and these are passband flux derivatives
over effective temperature and gravity as well as limb darkening, included in the terms $D_{1,\ell}^x$ and $D_{3,\ell}^x$, respectively.
Here, we rely on two grids of stellar atmosphere models: LTE \citep{kurucz} and non-LTE \citep{hubeny} and consider different values
of the atmospheric metallicity, [m/H], and the microturbulent velocity, $\xi_t$.
The second input is connected with pulsational properties and this is the so-called nonadiabatic complex parameter $f$
describing a ratio of the amplitude of the radiative flux perturbation to the radial displacement at the photosphere level
(e.g. Daszy\'nska-Daszkiewicz, Dziembowski, Pamyatnykh 2003).The $f$-parameter is incorporated in the temperature term, $D_{1,\ell}^x$.

A comprehensive overview of possible sources of uncertainties in the theoretical values of the photometric amplitudes and phases
of early B-type pulsators has been studied in details by \citet{dsz2011}.

The value of $f$ for a given mode can be either obtained from the linear nonadiabatic computations
of stellar pulsations (Cugier, Dziembowski, Pamyatnykh 1994) or determined from observations (Daszy\'nska-Daszkiewicz et al. 2003, 2005).
Here, we applied the two approaches.

To identify the degree $\ell$ of pulsational modes of 12 Lac using the theoretical $f$-parameter, we compared
the theoretical and observational values of the photometric amplitude ratios and phase differences in various pairs of passbands.
To this aim, we considered models from the error box corresponding to masses from 10 to $12M_\odot$
and used the nonadiabatic pulsational code of Dziembowski (1977). To check how our identification of $\ell$ is robust we
considered two values of metallicity, $Z=0.015$ and 0.010, two chemical mixtures, AGSS09 and GN93 (\citet{grevesse}),
two sources of the opacity tables, OPAL \citep{opal} and OP \citep{op}, and two sets of stellar model atmospheres, LTE and non-LTE.
The hydrogen abundance was fixed at $X=0.7$ and no overshooting from the convective core was allowed.
We considered the values of $\ell$ from 0 up to 6 and the frequency range appropriate to the observational values.

Using the theoretical values of $f$, we arrived at the following identification.
The dominant mode $\nu_1$ as well as $\nu_2$ are certainly dipoles ($\ell=1$).
Identification of $\ell$ for $\nu_3$ and $\nu_5$ is also unique; they are quadruple modes ($\ell=2$).
There is also no doubt about identification of $\ell$ for $\nu_4$; certainly this is a radial mode ($\ell=0$).
Identification of $\ell$ for the remaining frequencies is ambiguous: $\nu_6$ is most likely a dipole, but it may be also $\ell=2$ or 3;
$\nu_7$ can be $\ell=1,2,3,5$; for $\nu_8$ we were not able to associate any values of $\ell$;
$\nu_9$ is either $\ell=3$ or 1 and $\nu_{10}$ can be $\ell=1$ or 2 or 3.
Finally for the SPB like mode, $\nu_{\rm A}$, we got  $\ell=1$ or 6, but because of the strong averaging effects
which increase with the harmonic degree, the $\ell=6$ identification is much less probable.
These identifications are independent of the metallicity, opacity data,
mixture and atmosphere models. An influence of the different stellar parameters (within the error box) is also negligible.

\begin{table}
 \caption{The most probable identification of $\ell$ for the 12 Lac frequencies obtained from two methods described in the text.
In the last column, we put the values of $m$ determined by Desmet et al.\,(2009).}
 \label{identyfikacja}
 \begin{tabular}{@{}ccccc}
  \hline
     frequency   & $A_V$                        & \multicolumn{2}{c}{the most probable $\ell$}     & $m$ \\
    ~ [c/d] & [mmag] & theoret. $f$ & empir. $f$ & Desmet2009\\
  \hline \hline
  $\nu_1$=5.179034     & 38.1 &    1        &    1     & 1\\
  $\nu_2$=5.066346     & 16.0 &    1        &    1     & 0\\
  $\nu_3$=5.490167     & 11.1 &    2        &    2     & 1\\
  $\nu_4$=5.334357     & 10.0 &    0        &    0     & 0\\
  $\nu_5$=4.24062      &  3.6 &    2        &    2     & 0,1,2\\
  $\nu_6$=7.40705      &  2.0 &    1,2,3    &   1,2    &--\\
  $\nu_7$=5.30912      &  2.0 &    1,2,3,5  &    --    &--\\
  $\nu_8$=5.2162       &  1.3 &       ?     &    2     &--\\
  $\nu_9$=6.7023       &  1.3 &    3,1      &   3,1    &--\\
  $\nu_{10}$=5.8341    &  1.3 &    1,2,3    &   1,2    &--\\
  $\nu_{\rm A}$=0.35529 & 5.0 &    1,6      &    1,4   &--\\
  \hline
 \end{tabular}
\end{table}

In the second approach the mode degree $\ell$ is determined simultaneously with the empirical values of $f$
by fitting the theoretical values of the photometric amplitudes and phases to the observational counterparts
(Daszy\'nska-Daszkiewicz et al. 2003, 2005). To this end we considered stellar parameters from the center and edges of the error box.
We checked also the effects of adopted model atmospheres, i.e., effects of the microturbulent velocity, $\xi_t$,
metallicity, [m/H], as well as the non-LTE effects.
All the above parameters do not change identification of $\ell$ although a quality of the fit obtained with the LTE and non-LTE models
can differ significantly. Usually, with the non-LTE atmospheres we got the better goodness of the fit.
From this approach, we found a unique determination of $\ell$ for the six frequencies.
For the first five modes with the largest amplitudes results are consistent with the determinations
obtained with the theoretical values of the $f-$parameter.
In the case of some other frequencies we succeed in getting better constraints on $\ell$,
i.e., $\nu_6$ and $\nu_{10}$  can be $\ell=1$ or 2 and $\ell=3$ is excluded, $\nu_8$ has been identified uniquely as a quadruple.
The degree of $\nu_7$ is indeterminable because this frequency has not been detected in spectroscopy.
The slow mode $\nu_{\rm A}$ is now $\ell=1$ or 4.

In Table\,\ref{identyfikacja}, we list frequencies of 12 Lac and the most probable identification
of the degree $\ell$ from the two approaches. The last column contains the values of $m$ as determined by Desmet et al. (2009).
As we can see, identifications of $\ell$ from the two methods are consistent. In the case of $\nu_A$, the common identification is $\ell=1$
and this is the most likely degree. Moreover, the visibility of modes with $\ell=4$ and 6 is much lower.
Our identifications of $\ell$ agree with those obtained by Handler et al. (2006).

From the second approach, besides the degree $\ell$, we get the empirical value of $f$, which has a great asteroseismic potential,
and the intrinsic mode amplitude multiplied by the aspect factor, $\tilde\varepsilon=\varepsilon Y(i,0)$.
These quantities will be discussed in subsequent sections.

\subsection[]{Including effects of rotation}

Rotation causes that the $2\ell+1$ degeneracy is lifted what allows to determine the full geometry of pulsational modes,
i.e., the spherical harmonic degree, $\ell$, and azimuthal order, $m$. It also complicates mode identification because
the photometric amplitude ratios and phase differences become dependent on the inclination angle and rotational velocity.

In the case of low order p- and g-modes, the most important effects of rotation is mode coupling occurring
for close frequency modes with the same values of $m$ and $\ell$ differing by 2 (Soufi, Goupil \& Dziembowski, 1998).
The two modes are close if the difference between their frequencies
is of the order of the rotational frequency, i.e., $|\nu_1-\nu_2|\approx \nu_{\rm rot}$.
Then, the complex photometric amplitude due to a pulsational mode coupled by rotation is given as a sum of the photometric amplitudes
of modes contributing to the coupling (Daszy\'nska-Daszkiewicz et al. 2002)
\begin{equation}
$${\emph{\textbf{A}}}_\lambda^{\rm coup}(i)= \sum_k a_k {\cal A}_{\lambda,k} (i),$$
\end{equation}
where ${\cal A}(i)$ is given in Eq.\,1.
The coefficients $a_k$ results from linear perturbation theory of rotating stars and describe contributions
of the $\ell_k$-modes to the coupled mode.

According to \citet{desmet2009}, the surface equatorial rotational velocity of 12 Lac amounts to about 50 km/s.
Adopting the stellar radius of about 7 $R_\odot$ we get a surface rotational frequency of about $\nu_{\rm rot}=0.14$ c/d.
As has been shown by Daszy\'nska-Daszkiewicz et al. (2002), already rotation of this order can couple modes.

The dominant mode of 12 Lac, which has been identified as the $\ell=1$ mode, has the photometric amplitude
much higher than the other ones; its value of $A_V$ is about 2.5 larger than for $\nu_2$.
\textbf{Because usually, it is the radial mode which dominates, we checked whether $\nu_1$ could be the $\ell=0$ mode coupled with the $\ell=2$ mode by rotation.}
Pulsational models which satisfy the rotational coupling conditions for the frequency $\nu_1$ can be found
in a very wide range of stellar parameters within the 12 Lac error box.

\begin{figure*}
\begin{center}
 \includegraphics[clip,width=\textwidth,height=80mm]{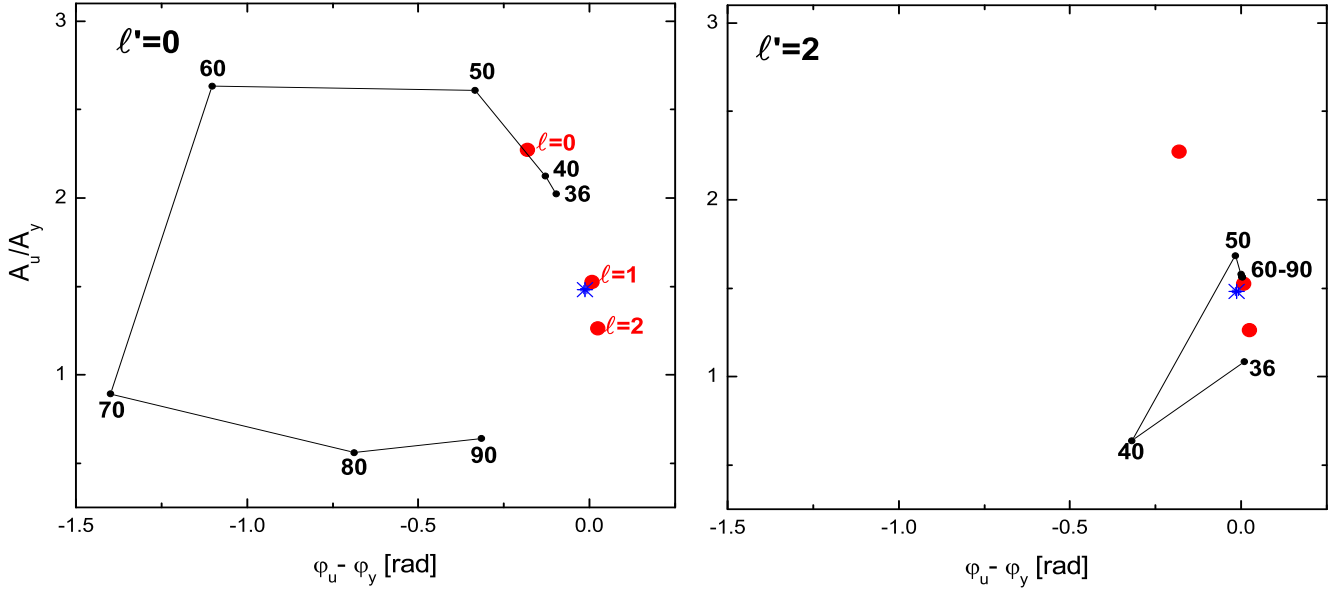}
 \caption{The diagnostic diagrams with Str\"omgren passbands $u$ and $y$, showing positions for coupled $\ell= 0, p_1$
and $\ell = 2, g_1$ modes excited in the model with $M= 9.43 M_\odot$, $\log T_{\rm eff}=4.3411$ and $\log L/L_\odot = 3.975$.
Big dots indicate positions of the pure modes with $\ell = 0, 1, 2$. The numbers at small dots are the values of the rotational
velocity in km/s. The values of the inclination angle result from the condition $V_{\rm rot}\sin i=36$ km/s.
An asterisk denotes an observational position of $\nu_1$ with errors.}
\label{coupl}
\end{center}
\end{figure*}

In Fig.\,\ref{coupl}, we show an example of a mode coupling between the $\ell=0,p_1$ and $\ell=2,g_1$ modes
with frequencies before the coupling of about 5.17 and 5.15 c/d, respectively.
The parameters of the model are:  $M= 9.43 M_\odot$, $\log T_{\rm eff}=4.3411$ and $\log L/L_\odot = 3.975$.
The value of $V_{\rm rot}\sin i=36$ km/s was kept constant and the values of $V_{\rm rot}$ are indicated at each point.
The pure modes with $\ell=$0, 1, 2 are marked as big dots. The observational position for the frequency $\nu_1$
is shown as an asterisk.
As we can see the best agreement is obtained if $\nu_1$ is considered as the pure $\ell=1$ mode. The same conclusion was reached
for other models of 12 Lac and other pairs of modes.
Therefore, in our further analysis we assume that $\nu_1$ is the pure $\ell=1$ mode.
\begin{figure*}
\begin{center}
 \includegraphics[clip,width=85mm,height=88mm]{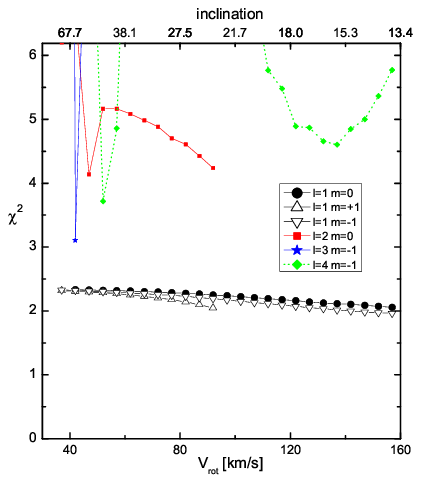}
 \includegraphics[clip,width=85mm,height=88mm]{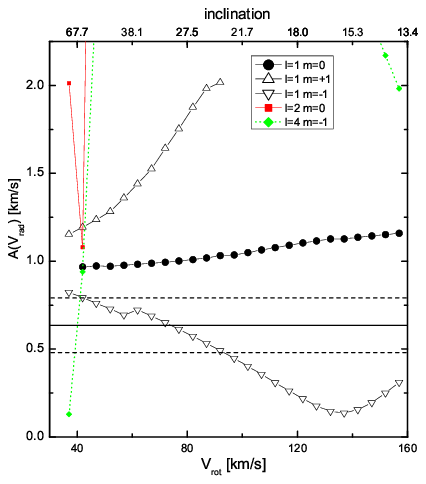}
\caption{Left panel: the values of $\chi^2$ for the $\nu_A$ frequency as a function of the rotational velocity.
Only unstable modes in the model with $M= 9.91 M_\odot$, $\log T_{\rm eff}=4.3546$ and $\log L/L_\odot = 4.039$
were included. The corresponding values of the inclination angle result from a condition $V_{\rm rot}\sin i=36$ km/s
and are indicated at the top $X$ axis. Right panel: the resulting amplitudes of the radial velocity variations
calculated for the optimal values of $\varepsilon$. The observational range of $A(V_{\rm rad})$ is marked as horizontal lines.}
\label{slow-chi}
\end{center}
\end{figure*}

In the case of high-order g-modes, whose frequencies are of the order of the rotational frequency, incorporation of the Coriolis force
is crucial. One way to account for these effects of rotation is the so-called traditional approximation (e. g. Lee \& Saio 1987, Townsend 2003).
In the case of the low frequency of 12 Lac, $\nu_A=0.35529$ c/d, the spin parameter, $s=2\nu_{\rm rot}/\nu$,
amounts to 0.8-1.0 depending on the model. At such value of $s$, the photometric amplitudes and phases can be changed.
The effects of rotation on photometric observables for high-order g-modes excited in the SPB star models were studied by Townsend (2003).
The formulas for the light and radial velocity variations, we use here, were given by Daszy\'nska-Daszkiewicz, Dziembowski \& Pamyatnykh (2007).
The semi-analytical form of the formula for the complex amplitude of the light variations due to pulsations in a high-order g-mode is as follows:
\begin{equation}
$${\emph{\textbf{A}}}_\lambda^{\rm trad}(i)= \sum_j \gamma_{\ell_j} {\cal A}_{\lambda,\ell_j} (i),$$
\end{equation}
where $\gamma_{\ell_j}$ are coefficients of the expansion of the Hough function into the series of the Legendre functions
(see Daszy\'nska-Daszkiewicz, Dziembowski \& Pamyatnykh, 2007) and ${\cal A}(i)$ is given in Eq.\,1.

In order to determine the angular numbers of $\nu_A$, we apply the same discriminant, $\chi^2$, as proposed by
Daszy\'nska-Daszkiewicz, Dziembowski \& Pamyatnykh (2008). The $\chi^2$ discriminant includes the best value of the intrinsic mode amplitude, $\varepsilon$,
to fit photometric amplitudes and phases in all passbands simultaneously. The value of $\chi^2$ is a function of both the inclination angle and rotational velocity.

In the left panel of Fig.\,3, we plot the dependence of $\chi^2$ on the rotational velocity for unstable modes with frequencies close to $\nu_A$.
Again the value of $V_{\rm rot}\sin i=36$ km/s was kept constant.
We show an example for model with the following parameters: $M= 9.91 M_\odot$, $\log T_{\rm eff}=4.3546$ and $\log L/L_\odot = 4.039$.
Only modes which give $\chi^2<6$ were shown. As we can see, all dipole modes are acceptable. For a possible discrimination of the azimuthal order, $m$,
in the right panel of Fig.\,3, we plotted the amplitude of the radial velocity variations obtained for the best value of $\varepsilon$ as a function of $V_{\rm rot}$.
The horizontal lines mark the observed value of $A(V_{\rm rad})$ with errors. We can see that only the retrogarde mode ($\ell=1,~m=-1$)
has the radial velocity amplitude within the observational range. This identification does not change if other models of 12 Lac are considered.
Thus, we assume that $\nu_A$ is a dipole retrograde mode.

\subsection{Radial orders of pulsational modes}

\textbf{Our assignment of the radial order, $n$, has been done within the zero-rotation approximation.
For non-axisymmetric frequencies with the identified azimuthal order, $m$, we estimated the centroid frequency, $\nu_{n \ell}$, from the following equation
\begin{equation}
\nu_{n \ell m}=\nu_{n \ell}+(1-C_{n \ell})m\nu_{\rm rot},
\end{equation}
where $C_{n,\ell}$ is the Ledoux constant and $\nu_{\rm rot}$ is the rotational frequency.}

In the HR diagram (Fig.\,1), we plotted the lines of constant period $P=0.18746$ d, corresponding to the radial mode $\nu_4$,
assuming that $\nu_4$ is the fundamental ($n=1$) or first overtone mode ($n=2$). Two values of metallicity, $Z=0.01$ and 0.015, were considered.
Both lines are within the error box, but the $n=2$ line is cut near the middle, for the lower metallicity models.
It is caused by a lack of the main sequence models reproducing $\nu_4$ with masses lower than about 11 $M_\odot$ for $Z=0.01$.

\begin{figure}
\begin{center}
\includegraphics[clip,width=75mm,height=55mm]{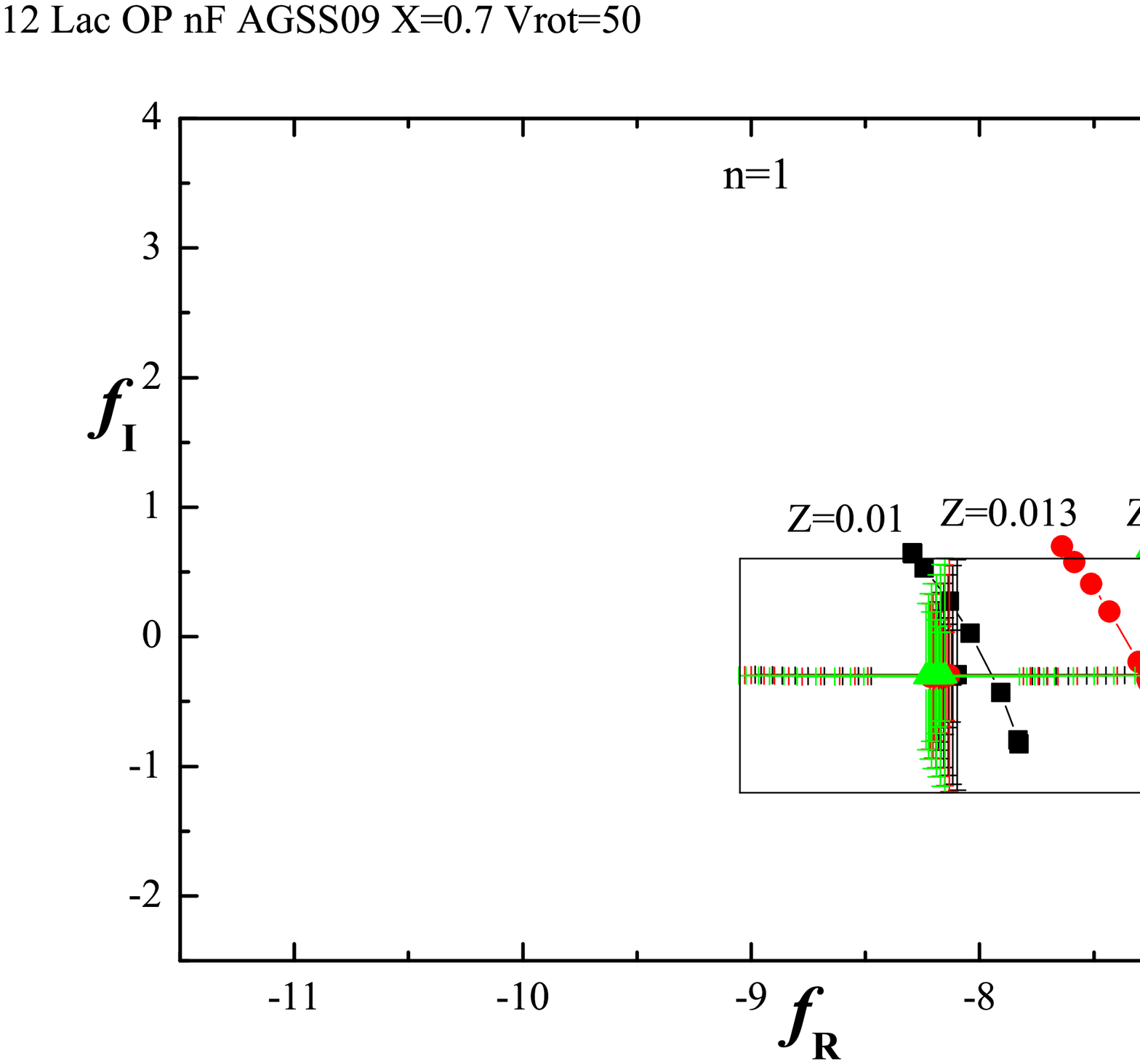}
\includegraphics[clip,width=75mm,height=55mm]{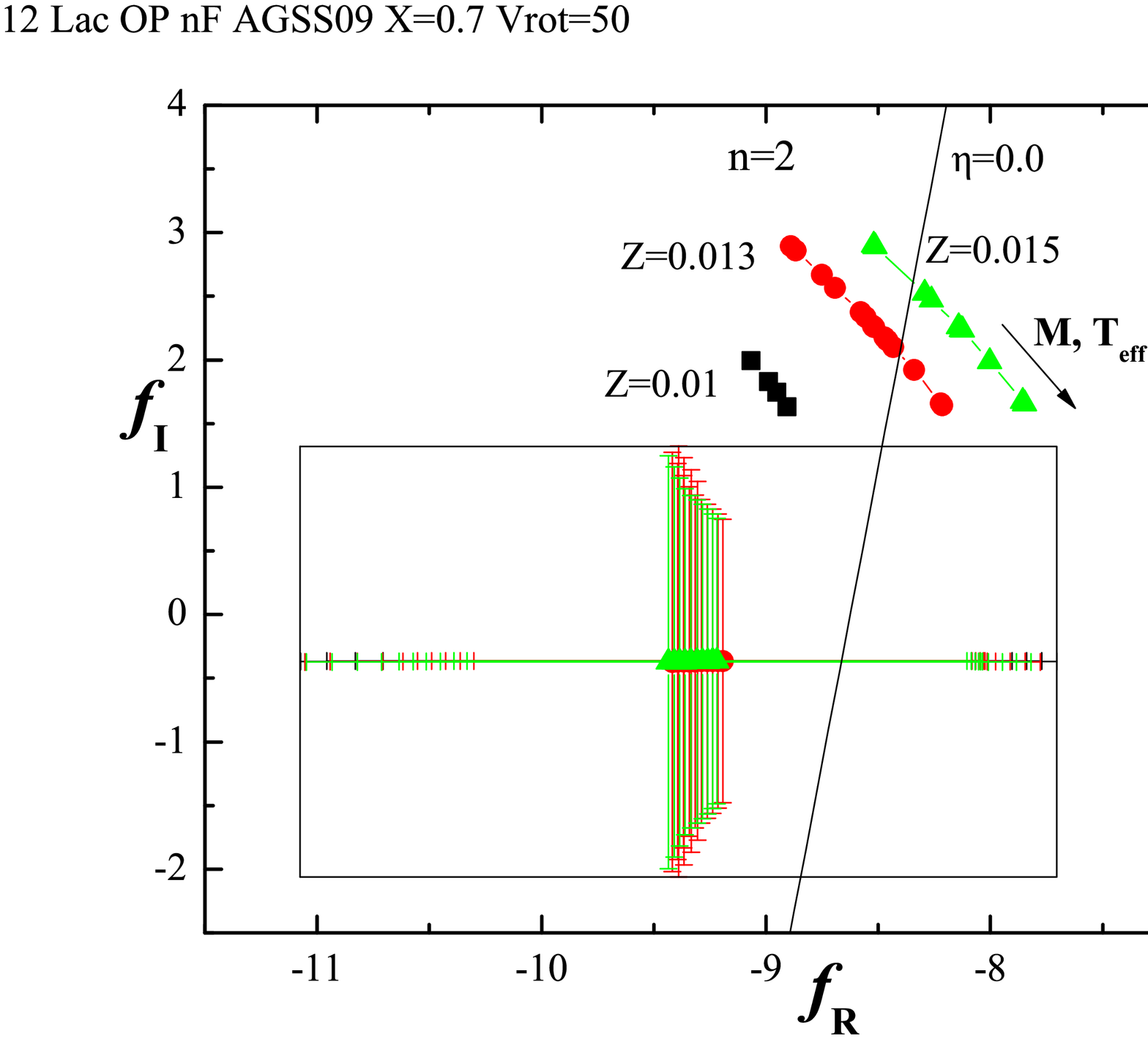}
\caption{Comparison of the empirical (boxes) and theoretical (lines with symbols) values of $f$ for the radial mode, $\nu_4$, on the complex plane.
The upper and bottom panels correspond to the hypothesis of the fundamental and first overtone radial mode, respectively.
The theoretical values of the $f$-parameter were calculated with the OP tables for three values of metallicity, $Z=0.01$, 0.015 and 0.02.
All models reproduce the frequency $\nu_4$ and are inside the error box of 12 Lac. The thin, black line in the bottom panel indicates the instability border.
Models to the right of this line are unstable.
In the upper panel, all models are unstable. The arrows indicates the increasing masses and effective temperatures.}
\label{fr-fi}
\end{center}
\end{figure}

In order to identify the radial order, $n$, of $\nu_4$, we compared the empirical and theoretical values of the nonadiabatic $f$-parameter
(Daszy\'nska-Daszkiewicz, Dziembowski \& Pamyatnykh, 2003, 2005). We chose models which are located on lines of constant period and are inside
the observational error box of 12 Lac. For these models, we computed the theoretical values of  $f$ for the frequency $\nu_4$
and compare them with the empirical counterparts. The results are presented in Fig.\,\ref{fr-fi}, where
the real part of the $f$-parameter, $f_{\rm R}$, is plotted against its imaginary part, $f_{\rm I}$.
The empirical values are marked as a box and the theoretical ones as lines with symbols corresponding to the three values of metallicity: $Z=$0.010, 0.013, 0.015.
All models were computed with the OP opacities and no overshooting from the convective core.
In the upper panel of Fig.\,\ref{fr-fi}, we assumed that $\nu_4$ is the fundamental mode, while in the bottom panel - the first overtone.

As we can see, the agreement is achieved if $\nu_4$ is the radial fundamental mode. Moreover, all these models are unstable,
while in the case of the first overtone, only models to the right of the thin, black line labeled as $\eta=0.0$ are unstable.
We need quite high metallicity and large masses to make $\nu_4$ unstable as a first overtone.
With the OPAL data, identification of $n$ is not so unambiguous but a requirement of instability and lower values of $Z$
make the first overtone also much less likely.
\textbf{The effect of the overshooting parameter, $\alpha_{\rm ov}$, on identification of $n$ is completely negligible.}
Thus, we adopted that $\nu_4$ is the radial fundamental mode in our further analysis.
Our identification of $n$ for the frequency $\nu_4$ confirms that of Dziembowski \& Pamyatnykh (2008)

Following \citet{desmet2009}, we assumed that the dipole mode $\nu_2$ is axisymmetric ($m=0$).
A survey of pulsational models with different masses, $M$, metallicities, $Z$, and core overshooting parameters, $\alpha_{\rm ov}$,
showed, that if $\nu_4$ is the radial fundamental mode, then $\nu_2$ can be only the g$_1$ mode.
To account for the rotational splitting of $\nu_1$ and $\nu_2$, which belong to the $\ell=1$, g$_1$ triplet,
Dziembowski\& Pamyatnykh (2008) suggested a nonuniform rotation with an increase of its rate between the envelope and core.
Therefore, we considered rotational splitting according to the two values of the rotational velocity: 50 and 80 km/s. The values of the Ledoux constant were
appropriate for models of 12 Lac.
Identification of $n$ for the first five modes is consistent for these two values of $V_{\rm rot}$.
The $\nu_1$ dipole mode was identified as the prograde mode ($m=1$) and it is the second component of the g$_1$ triplet.
The $\nu_3$ quadruple mode is the prograde mode ($m=1$) and it is a component of the g$_1$ quintuplet.
For the $\nu_5$ quadruple mode, \citet{desmet2009} were able to derive only that $m\ge 0$ what for most models
fit to the g$_2$ quintuplet.

As we have shown in Section\,3, $\nu_A$ is the dipole retrograde mode ($\ell=1,~m=-1$).
Identification of the radial order $n$ for low frequency modes is very sensitive to the adopted value of the rotational velocity.
Assuming $V_{\rm rot}$=50 km/s we got that $\nu_A$ is a $g_{15}$ or $g_{16}$ mode and for $V_{\rm rot}$=80 km/s -  $g_{13}$ or $g_{14}$ are allowed.

The remaining frequencies of 12 Lac do not have unambiguous identifications of the angular numbers so their radial orders cannot be determined as well.

\begin{figure}
 \begin{center}
 \includegraphics[clip,width=85mm,height=72mm]{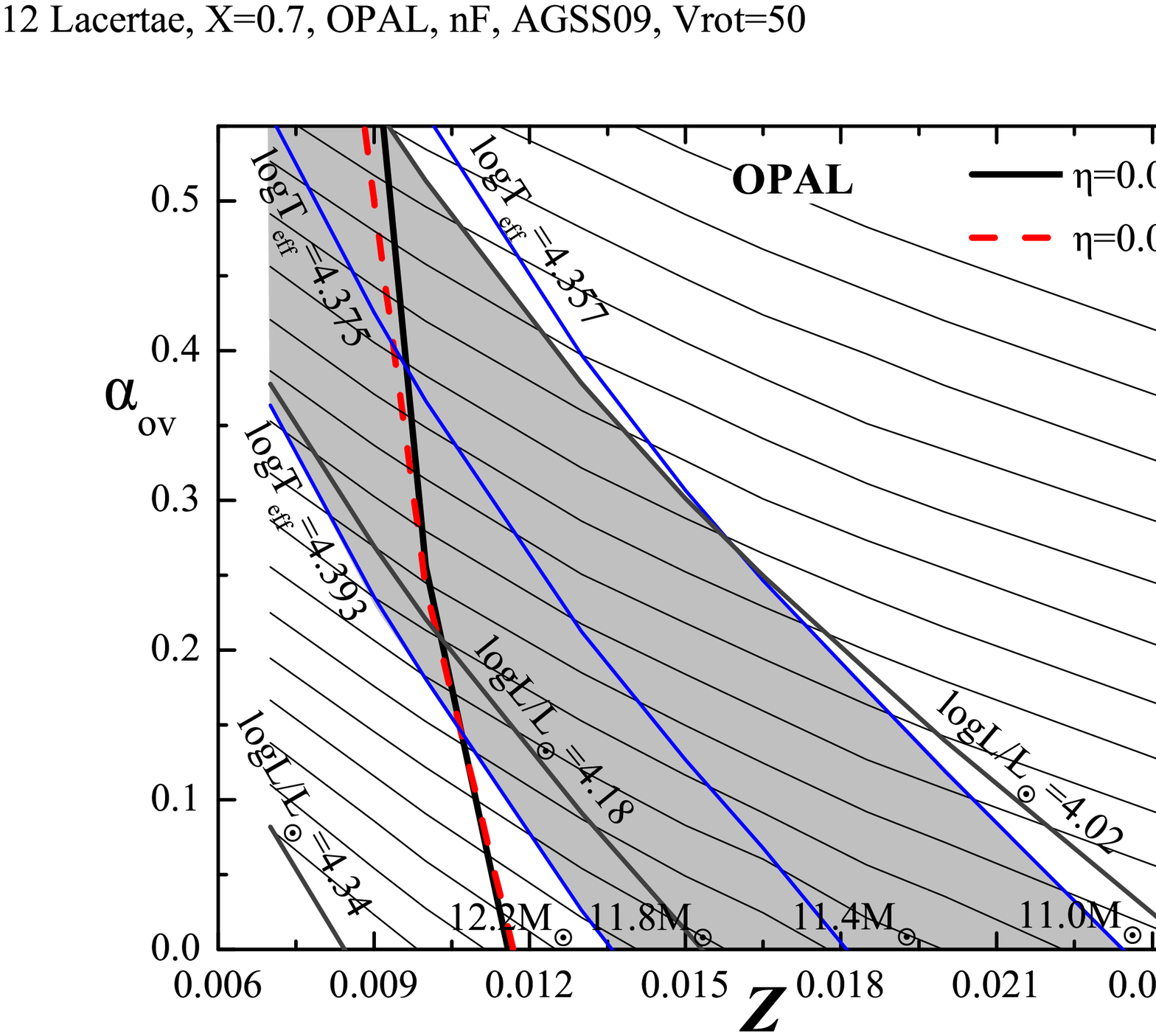}
 \caption{The overshooting parameter, $\alpha_{\rm{ov}}$, as a function of metallicity, $Z$, for seismic models of 12 Lac found from fitting the frequencies
 $\nu_{4}$ and $\nu_2$ (the modes $\ell = 0$, p$_1$ and $\ell = 1$, g$_1$, respectively) for the hydrogen abundance of $X=0.7$ and the OPAL opacities.
 There are plotted also lines of constant mass (thin, black), effective temperature (black), luminosity (grey) and instability borders for the radial p$_1$ mode
 (thick, solid) and dipole g$_1$ mode (thick, dotted). Models to the right of these lines are unstable. The grey area indicates models inside the error box.}
\label{Z-AO-OPAL}
\end{center}
\end{figure}

\begin{figure*}
 \begin{center}
 \includegraphics[clip,width=84mm,height=70mm]{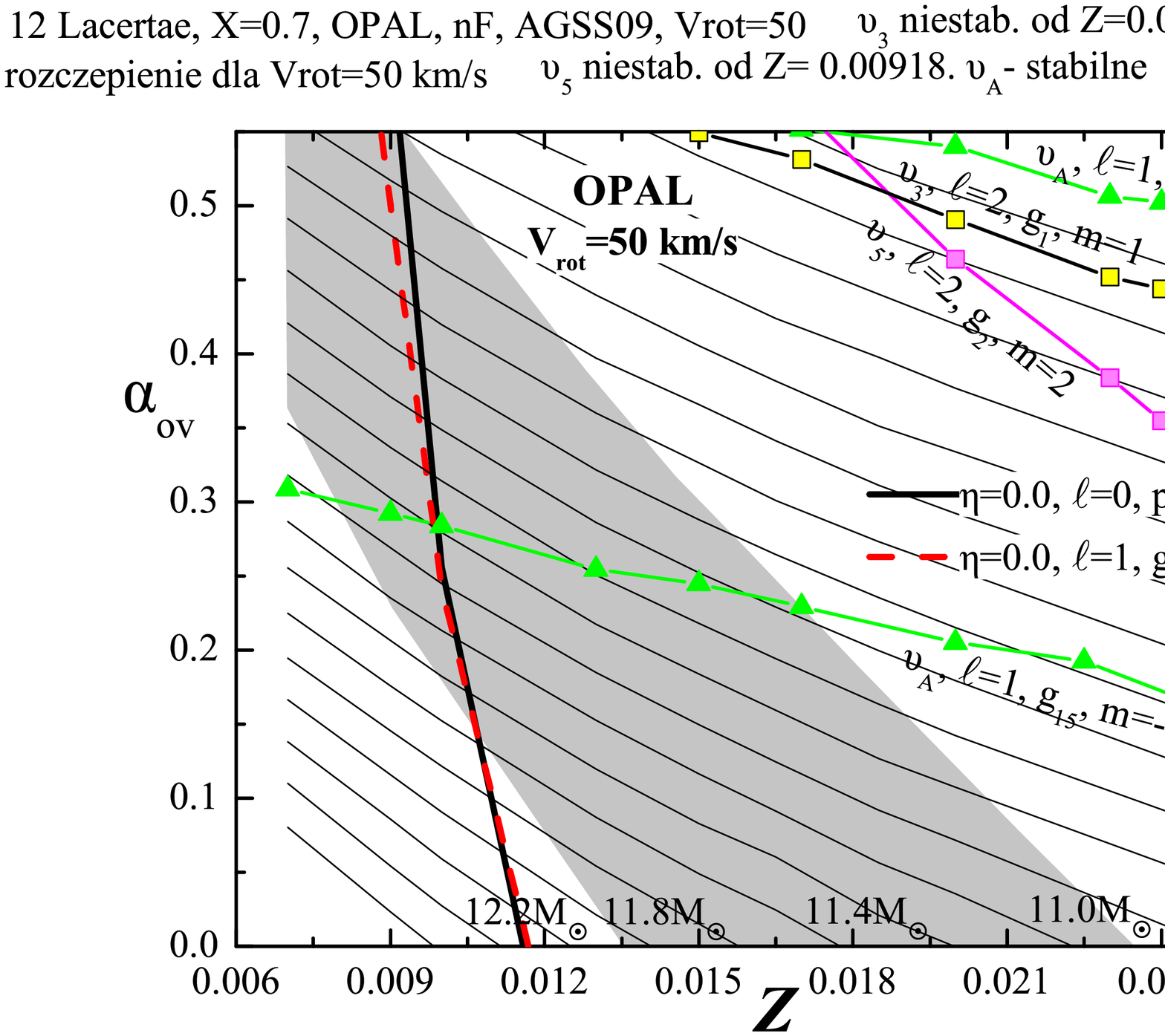}
 \includegraphics[clip,width=84mm,height=70mm]{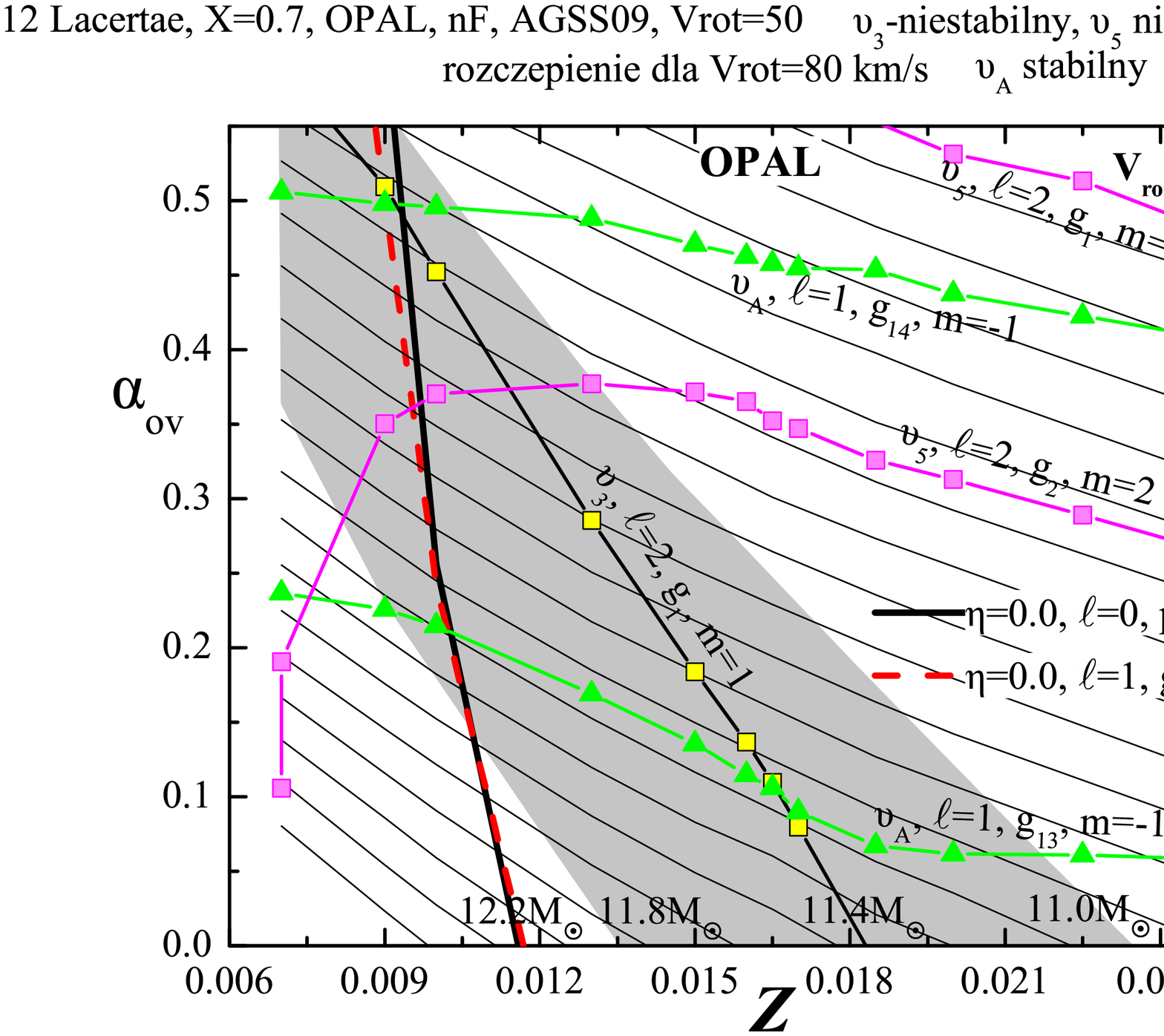}
 \caption{The same as in Fig.\,\ref{Z-AO-OPAL} but we added lines of models fitting the third frequency: $\nu_3$ or $\nu_5$ or $\nu_{A}$. The left and right panels correspond to models computed with the rotational splitting for $V_{\rm rot}=$ 50 and 80 km/s, respectively. The grey area indicates the observational error box.}
\label{Z-AO-OPAL-3}
\end{center}
\end{figure*}

\section{Complex asteroseismology}

Our seismic modeling can be divided into two parts.
First, we computed pulsational models which fit two well identified axisymmetric pulsational frequencies. In the case of 12 Lac, these are $\nu_4$ and $\nu_2$,
identified as the radial fundamental and dipole g$_1$ modes, respectively. Then, the theoretical values of the complex nonadiabatic parameter $f$ corresponding
to these two frequencies were compared with the empirical counterparts. We termed this approach {\it complex seismic model}.
Here, we construct complex seismic models for two sets of the opacity data (OPAL and OP), two sets of the model atmospheres (LTE and non-LTE),
and two sets of the chemical mixture (AGSS09 and the one derived for 12 Lac by \cite{morel2006}).

\begin{figure*}
\begin{center}
\includegraphics[clip,width=83mm,height=70mm]{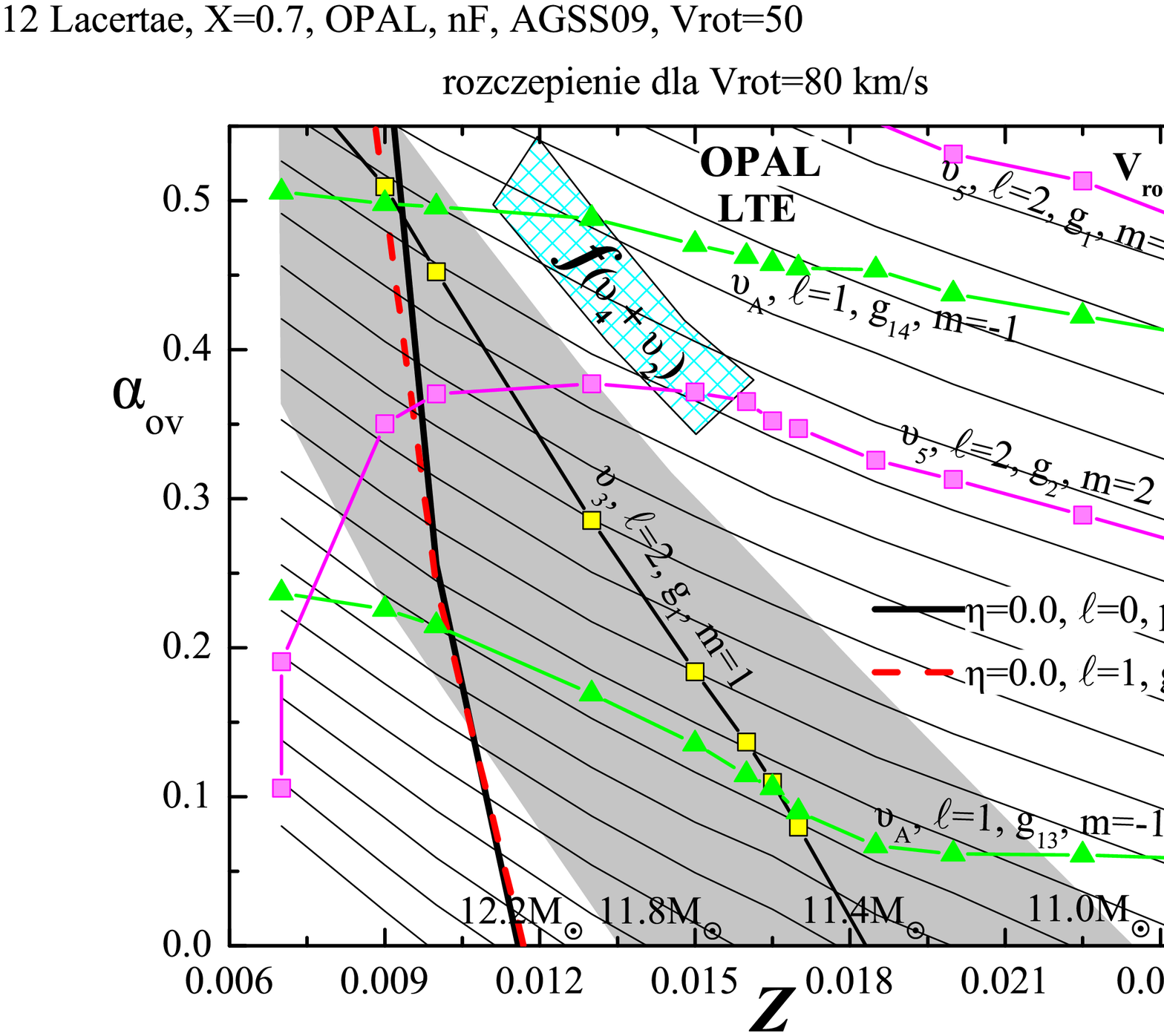}
\includegraphics[clip,width=83mm,height=70mm]{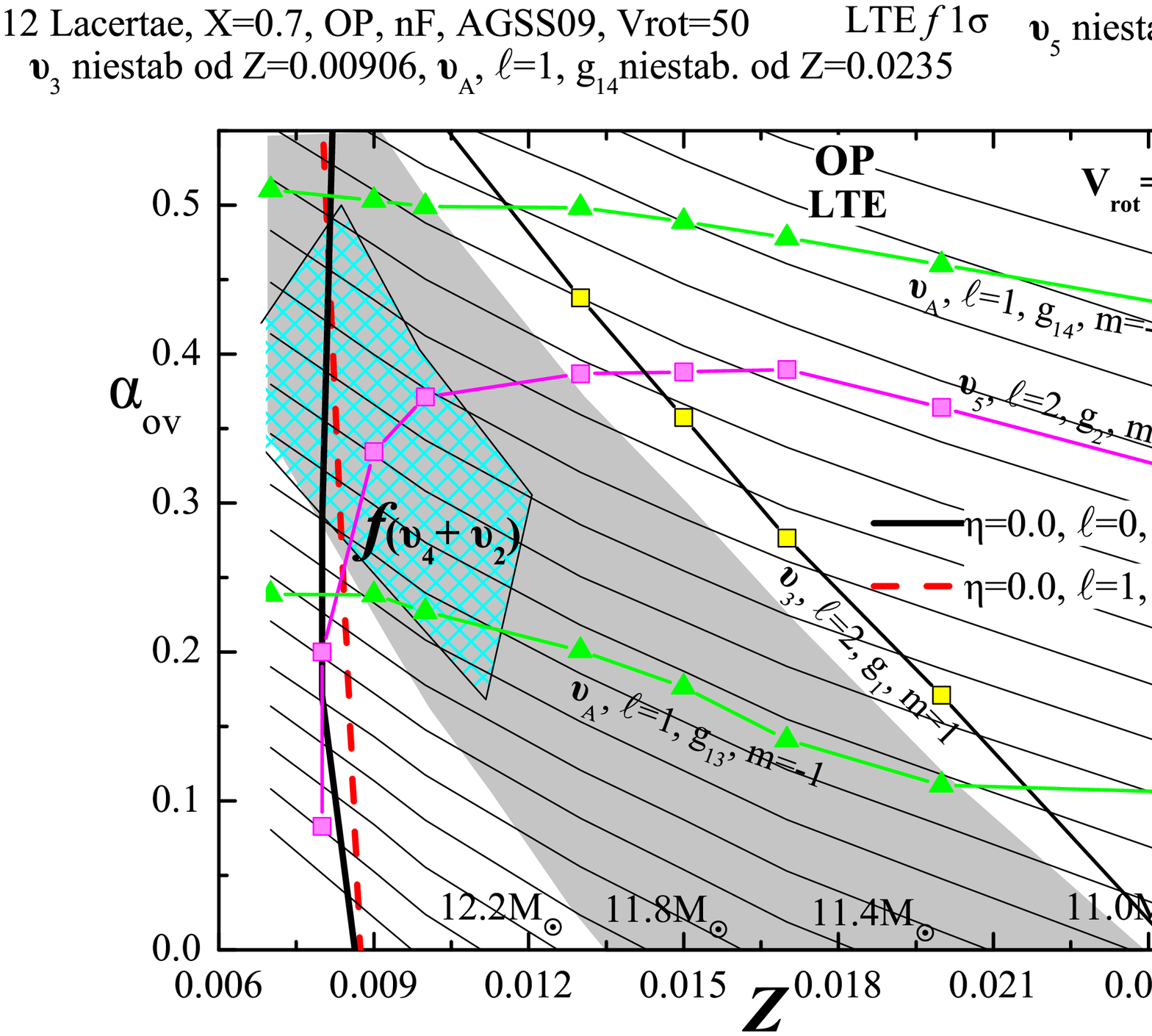}
 \caption{The same as in the right panel of Fig.\,\ref{Z-AO-OPAL-3}, but we marked models fitting the empirical values of the nonadiabatic $f$-parameter (hatched areas)
  of both the radial fundamental mode and dipole g$_1$ mode (labeled as $f(\nu_4 + \nu_2$). In the left panel, we used the OPAL data and in the right panel the OP
  opacities. In both cases we assumed the AGSS09 mixture and the rotational splitting for $V_{\rm rot}=80$ km/s. The LTE model atmospheres with the microturbulent
  velocity of $\xi_t=2$ km/s were used.}
\label{Z-AO-OPAL-OP-f}
\end{center}
\end{figure*}

\subsection{Fitting frequencies}

Fitting two, well-identified frequencies, $\nu_4$ and $\nu_2$, for an assumed hydrogen abundance, $X=0.7$,
allow to relate the mass, $M$, and the overshooting parameter, $\alpha_{\rm{ov}}$, of seismic models with their metallicity, $Z$.
These relations are presented in Fig.\,\ref{Z-AO-OPAL}, where the overshooting parameter is plotted versus metallicity.
Models with constant masses are marked with black, thin lines. Thick, solid line indicates instability border for the radial fundamental mode,
while thick, dotted line - for the dipole g$_1$ mode. The instability borders of these two modes almost overlap and cover a very narrow range
of metallicity, i.e., $Z\in (0.0105- 0.0115)$. Models to the right of these lines are unstable. We marked also the lines of constant effective temperature
and luminosity. The showed values of $\log{T_{\rm{eff}}}$ and $\log{L/L_{\odot}}$ correspond to the center and edges of the observational error box.

As we can see, there are a lot of seismic models of 12 Lac fitting two frequencies, $\nu_4$ and $\nu_2$. An accuracy of this fitting amounts to $10^{-6}$ c/d.
In general, models with a given value of the core overshooting have smaller masses, if they have larger metallicity. On the other hand, models with constant mass
have increasing metallicity with decreasing overshooting. These results are similar to those obtained for $\gamma$ Peg
(Walczak \& Daszy\'nska-Daszkiewicz 2010, Walczak et al. 2013).
The number of these seismic models can be limited if we take into account only unstable models and inside the observational error box (grey area).
This slightly narrow the metallicity parameter, $Z$, which should be from about 0.010 up to 0.024. Unfortunately, we are not able to constrain
the overshooting parameter, $\alpha_{\rm{ov}}$. There are seismic models with no overshooting from the convective core as well as with
$\alpha_{\rm{ov}}\sim0.5$.
\begin{figure*}
\begin{center}
 \includegraphics[clip,width=83mm,height=70mm]{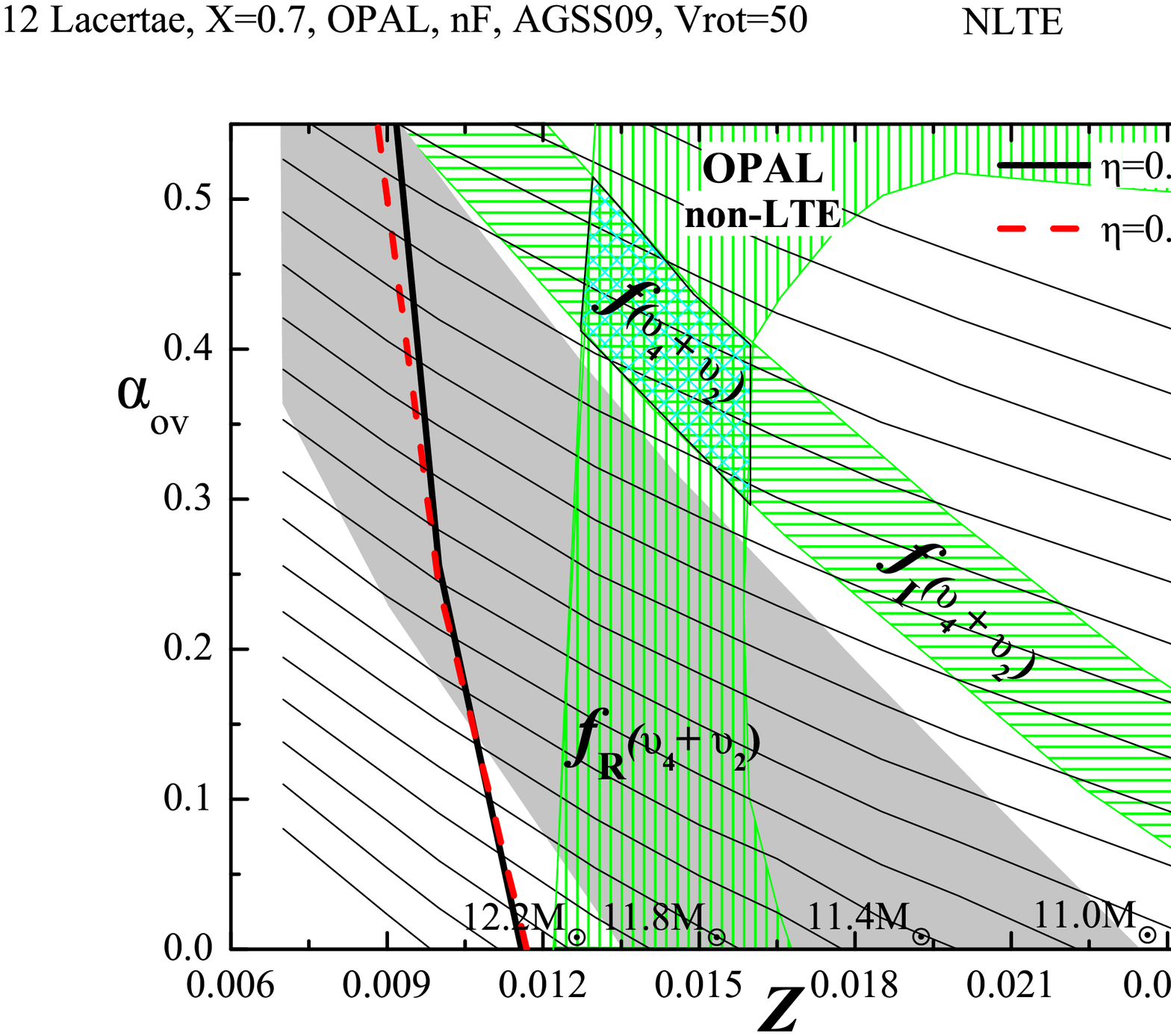}
 \includegraphics[clip,width=83mm,height=70mm]{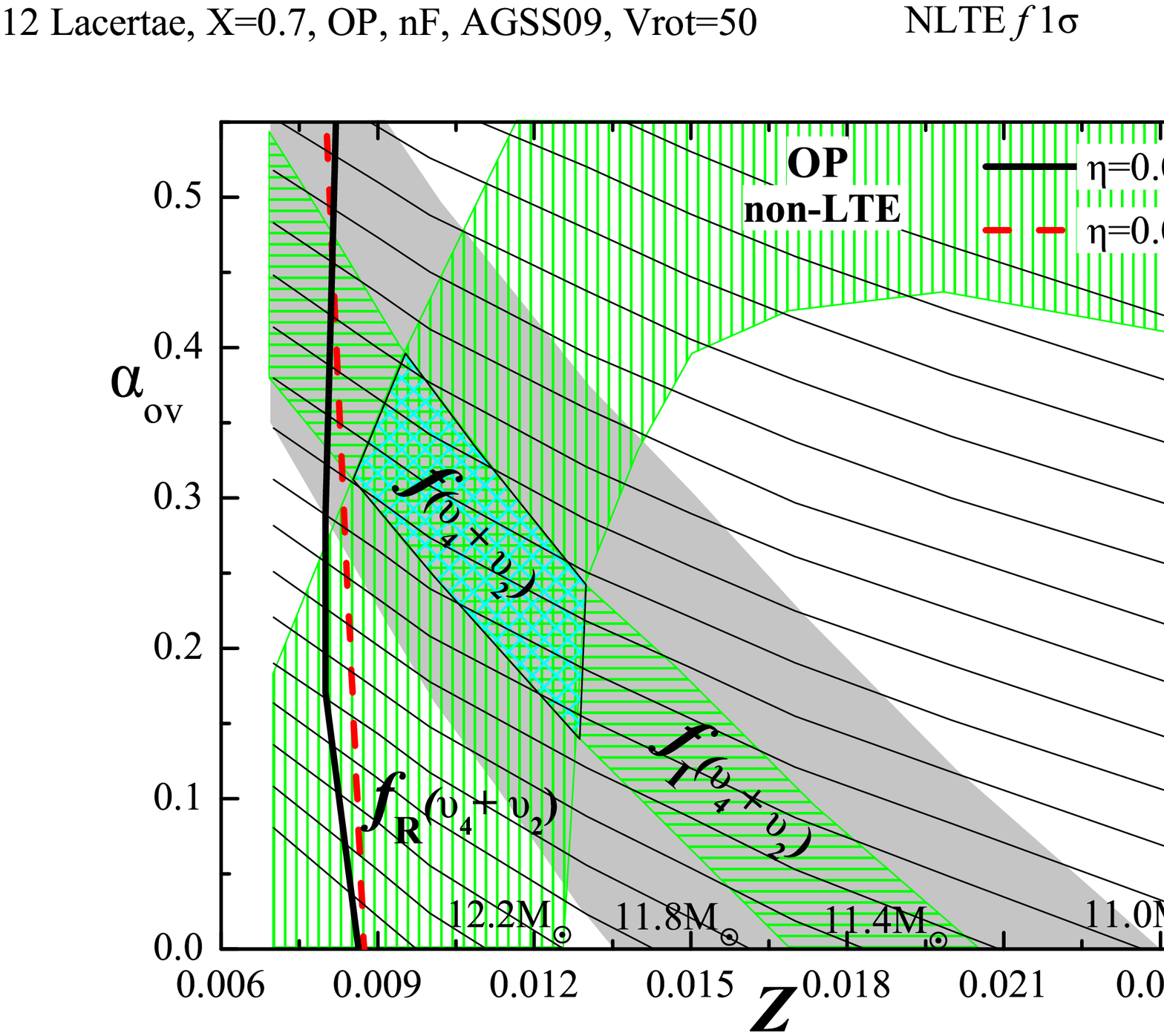}
 \caption{The same as in  Fig.\,\ref{Z-AO-OPAL-OP-f}, but we used the non-LTE model atmospheres in order to derived the empirical values of the $f$-parameter.
  There are also shown models which fit separately the real (vertical bars) and imaginary (horizontal bars) parts of $f$.}
\label{Z-AO-OPAL-OP-NLTE}
\end{center}
\end{figure*}

In the next step, we will check whether these models reproduce other well identified frequencies, i.e.,
$\nu_3=5.490167$ c/d - a quadruple prograde mode ($\ell=2,~m=1$), $\nu_5=4.24062$ c/d - a quadruple prograde mode ($\ell=2,~m>0$),
and $\nu_A=0.35529$ c/d - the dipole retrograde mode ($\ell=1,~m=-1$). The centroid values of these frequencies were calculated according to Eq.\,(12)
for two values of the rotational velocity, 50 and 80 km/s.
Lines of models fitting additionally one of these three frequencies are applied in Fig.\,\ref{Z-AO-OPAL-3}. The left and right panels include models
computed with the rotational splitting for $V_{\rm rot}=$ 50 and 80 km/s, respectively. The accuracy of the fitting of these three frequencies ranges
from $10^{-5}$ to $5\cdot 10^{-4}$ c/d.

As we can see, models computed with $V_{\rm rot}=50$ km/s reproduce $\nu_3$, $\nu_5$ and $\nu_A$ as a g$_{16}$ mode, only
for high values of the overshooting parameter and quite high metallicites. Models fitting $\nu_A$ as a g$_{15}$ mode are nearly
on the horizontal line and give constraints on overshooting: $\alpha_{\rm ov}\in (0.15,0.31)$. Here, $\nu_5$ can be only the quadruple sectoral mode
($\ell=2,~m=+2$).
In the case of models computed with $V_{\rm rot}=80$ km/s, we obtained lower values of the overshooting parameter.
The frequency $\nu_A$ has now lower radial order: $n=13$ for $\alpha_{\rm ov}\in (0.05,0.24)$ and $n=14$ for $\alpha_{\rm ov}\in (0.4,0.5)$
The frequency $\nu_5$ can be either $m=+2$ for $\alpha_{\rm ov}\in (0.1,0.38)$ or $m=+1$ for $\alpha_{\rm ov}\approx 0.5$ and $Z>0.018$.

It is worth to notice that models fitting additionally the frequency $\nu_A$ are usually on the nearly horizontal lines, i.e., they have
nearly constant overshooting regardless of the metallicity. Thus, the high-order g-modes could be potentially good indicators of $\alpha_{\rm ov}$
but proper identification of $n$ is crucial. A similar behaviour can be seen in the case of models fitting additionally $\nu_5$ except for the small
values of metallicity, $Z<0.01$. On the other hand the line of models fitting additionally the frequency $\nu_3$ corresponding
to $(\ell=2,m=+1, g_1)$, has the largest slope, i.e., the largest dependence of $\alpha_{\rm ov}$ on $Z$. In general, models fitting pressure modes
form the slopped lines on the the $\alpha_{\rm ov}~vs.~Z$ plane whereas models fitting gravity and mixed modes - nearly horizontal lines.

As we can see from the right panel of Fig.\,\ref{Z-AO-OPAL-3}, there are a few intersections of two lines which means that we have models fitting
four frequencies simultaneously: ($\nu_4, \nu_2, \nu_3, \nu_A$) or ($\nu_4, \nu_2, \nu_5, \nu_A$) or ($\nu_4, \nu_2, \nu_3, \nu_5)$.

\subsection{Fitting the $f$-parameters}

Each pulsational frequency is associated with the non-adiabatic complex parameter $f$ which gives the ratio of the bolometric flux changes
to the radial displacement. This quantity is very sensitive to properties of subphotospheric layers where mode driving occurs. In the case of B-type pulsators,
$f$ shows a strong dependence on metallicity and opacity data. Thus, a comparison of the theoretical and empirical values of $f$ may yield constraints
on these quantities (Daszy\'nska-Daszkiewicz, Dziembowski \& Pamyatnykh, 2005).

In a set of seismic models fitting the frequencies $\nu_4$ and $\nu_2$, we tried to find those which reproduce
also both values of the $f$-parameter within the observational errors.
Such models exist and are marked as hatched areas in Fig.\,\ref{Z-AO-OPAL-OP-f} on the $Z$ $vs.$ $\alpha_{\rm{ov}}$ planes and are labeled as $f(\nu_4+\nu_2)$.
The left and right panels correspond to models computed with the OPAL and OP data, respectively.
To derive the empirical values of $f$, we used the LTE model atmospheres with the microturbulent velocity of $\xi_t=2$ km/s \citep{kurucz}.

\begin{figure*}
\begin{center}
 \includegraphics[clip,width=83mm,height=70mm]{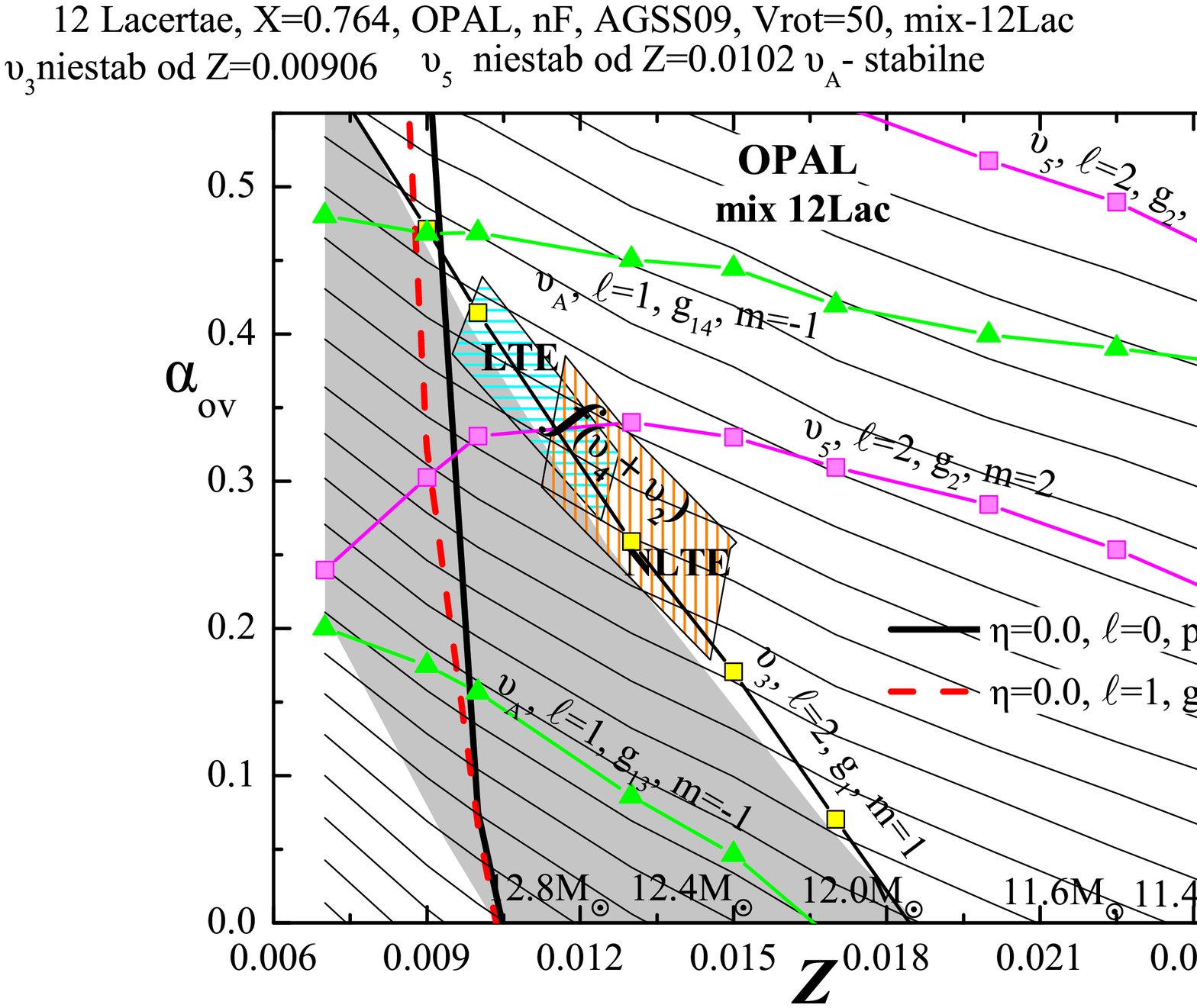}
 \includegraphics[clip,width=83mm,height=70mm]{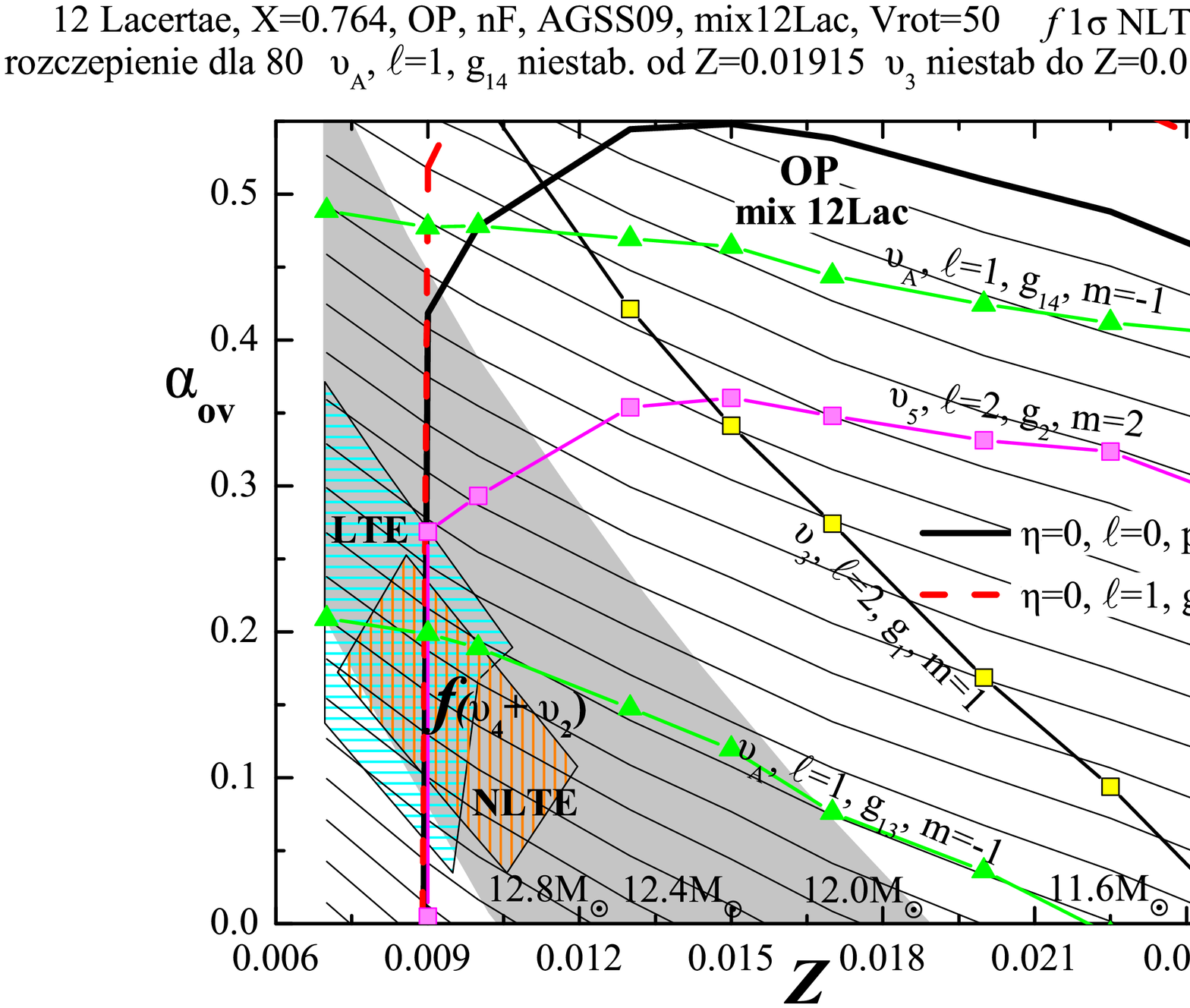}
 \caption{The same as in  Fig.\,\ref{Z-AO-OPAL} but for models computed for the 12 Lac mixture. We marked models reproducing the empirical values of the nonadiabatic $f$-parameter of both $\nu_4$ and $\nu_2$, obtained with the LTE (horizontal bars) and non-LTE (vertical bars) atmospheres.
In the left panel we used the OPAL data and in the right panel the OP tables.}
\label{Z-AO-OPAL-OP-mix}
\end{center}
\end{figure*}

As we can see, the hatched regions are quite small, what substantially reduces the number of models. The OPAL seismic models are slightly outside the
observational error box of 12 Lac. They are cooler and less luminous. They have also a high overshooting parameter, $\alpha_{\rm{ov}}\sim0.35-0.55$.
The OP seismic models have smaller metallicity and overshooting. In this case, the area $f(\nu_4+\nu_2)$ is larger and entirely inside the error box,
suggesting a preference for the OP tables.
In the case of the high-order g-mode, $\nu_A$, all OPAL models fitting this frequency reproduce also the empirical values of $f$.
The line ($\ell=1$, g$_{14}$) crosses the area $f(\nu_4+\nu_2)$ for $\alpha_{\rm ov}\approx 0.5$ and $Z\approx 0.014$.
Also almost all OP models fitting the frequency $\nu_A$ reproduce its value of $f$; except for models with the core overshooting of $\alpha_{\rm ov}\approx 0.45$
and metallicity of $Z\approx 0.020-0.025$.
In this case the line ($\ell=1$, g$_{13}$) crosses the area $f(\nu_4+\nu_2)$ for $\alpha_{\rm ov}\approx 0.2$ and $Z\approx 0.011$.
Thus with both opacity OP tables, we get seismic models reproducing the three frequencies $\nu_4,~\nu_2,~\nu_A$ and their $f$-parameters simultaneously,
but only the OP models are located inside the observational error box and have ,,reasonable'' values of the core overshooting.
In the case of $\nu_3$ and $\nu_5$, no seismic model reproduces their values of $f$ in the allowed range of parameters, neither with the OPAL and OP data.

Comparing the OPAL and OP seismic models, we can conclude that for a given $Z$ and $\alpha_{\rm{ov}}$, models calculated with these two opacity tables
have similar masses, effective temperatures and luminosities.
Instability borders for the radial fundamental and dipole g$_1$ modes are slightly larger for the OP tables.
The positions of lines representing models fitting additionally  $\nu_5$ or $\nu_A$ are quite similar, the difference is that
in the case of the OP models, there is no mode $\ell=2,m=+1, g_2$ for $\nu_5$. The biggest change is connected with $\nu_3$.
The line of the OP models fitting the frequency $\nu_3$ is shifted toward much larger metallicities.

The empirical values of the $f$-parameter depend slightly on the model atmospheres, so we decided to check this effect. In Fig.\,\ref{Z-AO-OPAL-OP-NLTE},
we show models fitting the empirical values of $f$ for the $\nu_4$ and $\nu_2$ frequencies derived with the non-LTE models of stellar
atmospheres (Lanz \& Hubeny, 2007). In the left and right panels we used the OPAL and OP opacity tables, respectively.
Additionally, we marked models fitting separately the real, $f_{R}$, (vertical bars) and imaginary, $f_{I}$, (horizontal bars) parts of the $f$-parameter.
As we can see, the overshooting parameter of models fitting $f_{I}$ decreases with increasing metallicity. In the case of models fitting the real part of $f$,
the dependence of $\alpha_{\rm{ov}}$ on $Z$ is different. First, for low values of $Z$, the overshooting parameter increases rapidly with metallicity and then,
for larger $Z$, it starts to decrease. With the LTE atmospheres, the shape of the regions fitting the real and imaginary part of $f$ is similar;
the main difference is in positions of the regions. The theoretical values of $f$ for $\nu_4$ and $\nu_2$ fit the empirical counterparts derived with
the non-LTE atmospheres for slightly larger metallicity and lower core overshooting.

\subsection{Effects of the chemical mixture and the opacity enhancement}

To evaluate the effect of the adopted chemical mixture, we performed our seismic modeling assuming the photospheric chemical element abundances of 12 Lac as derived by
\cite{morel2006}. In Fig.\,\ref{Z-AO-OPAL-OP-mix}, we show the results on the $\alpha_{\rm ov}~vs.~Z$ plane. The left and right panels correspond to computations
with the OPAL and OP data, respectively. In both cases, the empirical values of $f$ were determined adopting both the LTE (horizontal bars) and non-LTE (vertical bars) model atmospheres.

In a comparison with computations performed with AGSS09 and $X=0.7$, now the models fitting additionally the $f$-parameters of $\nu_4$ and $\nu_2$ have considerably
lower values of the core overshooting parameter, while their metallicity parameter was not changed much. The mean value of $\alpha_{\rm ov}$ for models fitting
the $f$-parameters with the 12 Lac mixture is approximately 0.13 lower. The lower overshooting occurs for both OP and OPAL opacities as well as
for the LTE and non-LTE atmospheres. The lines of constant masses and a position of a grey area corresponding to the error box were also moved towards lower
overshooting. These changes are caused mainly
by a higher hydrogen abundance, $X=0.764$, which we adopted after \cite{morel2006}. Abundances of others elements had no significant influence on the results.
As we can see lines of models fitting additionally one of the three frequencies $\nu_3$, $\nu_5$, $\nu_A$, also follow the general trend, i.e., a shift towards
slightly lower values of $\alpha_{\rm ov}$. In the case of the OPAL data, models fitting the $f$-parameters are still mostly
outside the error box. Amongst seismic models computed with the OP tables, there are those that fit also the frequency $\nu_A$ and its values of the $f$-parameter.
Thus, also with the 12 Lac mixture, there exist seismic models fitting these three frequencies and corresponding values of the $f$-parameter simultaneously.

\textbf{To summarize, we constructed seismic models with two sets of opacity tables, two sets of model atmospheres, two values of microturbulent velocity
and two mixtures of chemical elements. Now, the question is which case is ,,the best''. To pre-answer this question, in Table\,\ref{tab_model},
we listed the models with the minimum values of $\chi^2$ for each case. These models reproduce the centroid frequencies, $\nu_2$ and $\nu_4$,
and their empirical values of the $f$-parameter. The values of $\chi^2$ were obtained from fitting the photometric and radial velocity amplitudes and phases
when determining the empirical values of $f$ for $\nu_2$ and $\nu_4$ (see Sect.\,3.1 for the description of the method).}
\begin{table*}
\begin{center}
\caption{A list of the ,,best'' seismic models for each considered case. The basic stellar parameters are given in columns $2-6$.
The last two columns contains the values of $\chi^2$ as determined from fitting the photometric and radial velocity amplitudes and phases
for $\nu_2$ and $\nu_4$, respectively.}
\begin{tabular}{ccccccccc}
\hline

Model & $Z$	& $\alpha_{\rm ov}$	& $M$ [$M_{\odot}$]	&$\log T_{\rm eff}$ & $\log L/L_{\odot}$ & $\chi^2(\nu_2)$ & $\chi^2(\nu_4)$ \\
\hline
OPAL LTE $\xi_t=8$& 0.0135&0.45&9.3027&4.3463&3.9711&2.00&2.76\\
OPAL LTE $\xi_t=2$& 0.0135&0.40&9.5440&4.3513&3.9973&3.54&4.05\\
OPAL NLTE $\xi_t=2$&0.0160&0.30&9.8440&4.3501&4.0003&2.77&3.45\\
\hline
OPAL mix12Lac LTE &0.0100&0.41&10.289&4.3522&4.0217&3.49&3.84\\
OPAL mix12Lac NLTE&0.0125&0.25&10.922&4.3549&4.0469&4.54&6.58\\
\hline
OP LTE $\xi_t=2$ &0.0115&0.32&10.152&4.3693&4.0846&8.95&5.80\\
OP NLTE $\xi_t=2$&0.0130&0.24&10.462&4.3704&4.0962&6.20&5.32\\
\hline
OP mix12Lac LTE  &0.0100&0.21&11.494&4.3733&4.1322&10.98&6.46\\
OP mix12Lac NLTE &0.0115&0.12&11.899&4.3749&4.1464&7.62&5.11\\
\hline
\end{tabular}
\label{tab_model}
\end{center}
\end{table*}

\begin{figure*}
\begin{center}
 \includegraphics[clip,width=163mm,height=120mm]{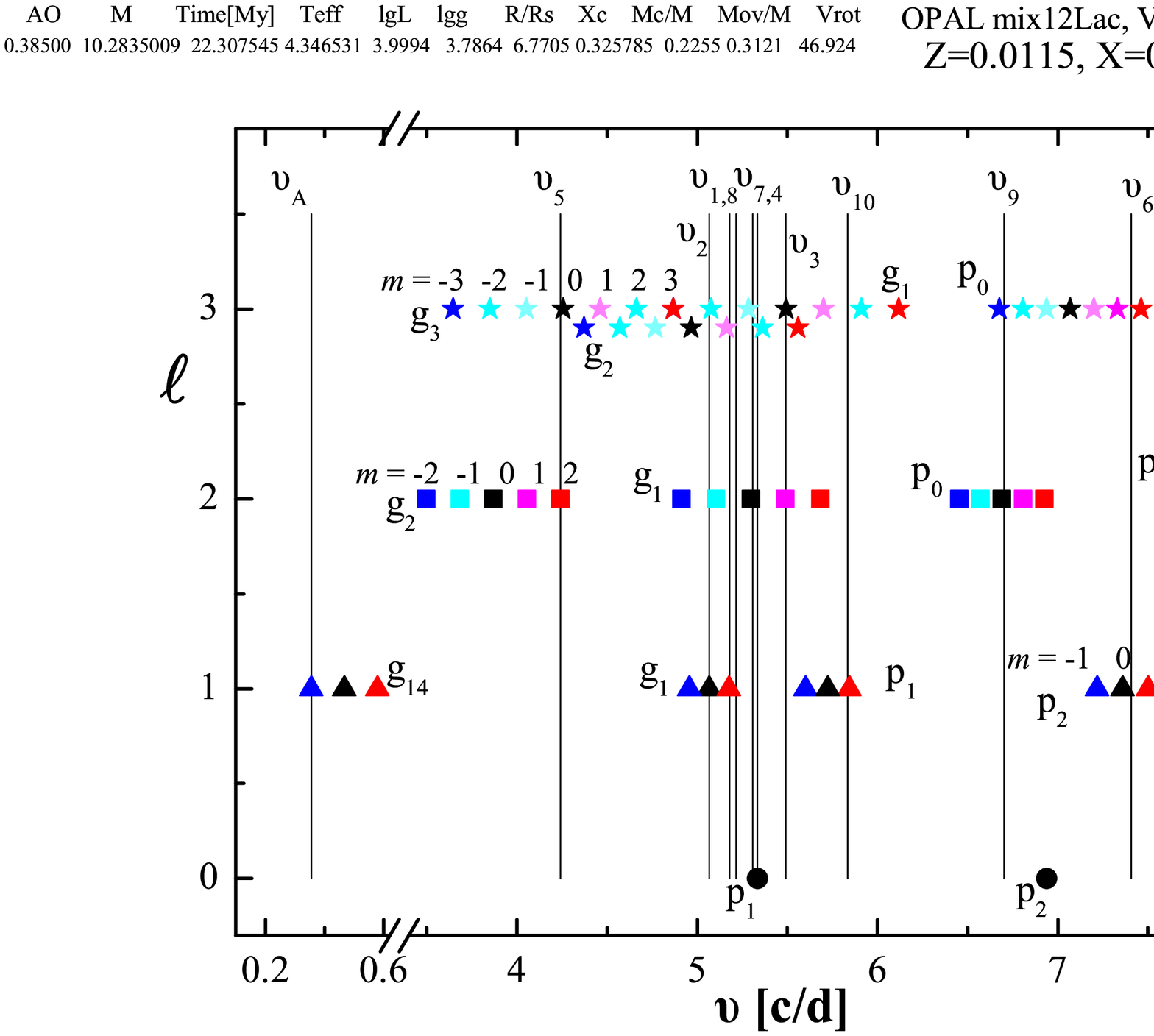}
 \caption{Comparison of the observed frequencies of 12 Lac (vertical lines) with theoretical ones (symbols) corresponding to MODEL-$6\nu$ fitting six frequencies:
 $\nu_1$, $\nu_2$, $\nu_3$, $\nu_4$, $\nu_5$ and $\nu_A$. The value of the rotational splitting was computed at $V_{\rm rot}=50$ km/s
 for p-mode frequencies and at $V_{\rm rot}=74.6$ km/s for g-mode frequencies. The mode degrees $\ell=0,1,2,3$ were considered.}
\label{6ni_model}
\end{center}
\end{figure*}

In the next step, we were trying to find a model fitting the first five frequencies, $\nu_1$, $\nu_2$, $\nu_4$, $\nu_3$, $\nu_5$, as well as $\nu_A$.
We did it by setting the value of the rotational velocity which gives an adequate splitting for $\nu_1$, $\nu_3$, $\nu_5$, $\nu_A$.
We obtained $V_{\rm rot}=$74.6 km/s what is consistent
with the value suggested by Dziembowski \& Pamyatnykh (2008) for the inner part of the star. In the case of these four mode a significant contribution
to the kinetic energy density comes from the g-mode propagating zone and their splitting is mainly determined by the inner layers which presumably rotate faster.
This rotational rate agrees also with the value of $V_{\rm rot}$ obtained from identification of the angular numbers of the g-mode frequency, $\nu_A$,
(cf. the right panel of Fig.\,3).
In Fig.\,\ref{6ni_model}, we compare the observed frequency spectrum of 12 Lac with the theoretical one corresponding to the model reproducing the six
frequencies. The model has the following parameters: $M=10.28 M_\odot$, $\log T_{\rm eff}=4.347$ and $\log L/L_\odot=4.00$,
and was computed for the metallicity $Z=0.0115$, overshooting parameter $\alpha_{\rm ov}=0.39$, OPAL tables and the 12 Lac mixture.
We called it MODEL-$6\nu$ and it is marked as an asterisk in the HR diagram (Fig.\,1).
The theoretical p-modes were split according to the rotational rate of $V_{\rm rot}=50$ km/s which corresponds to the outer layers.

As we can see in Fig.\,\ref{6ni_model}, the dipole prograde mode reproduces very well the dominant frequency, $\nu_1$.
The frequency $\nu_6$, identified as $\ell=1$ or 2 is quite close to the centroid of $\ell=1,$ p$_2$ and to $\ell=3,$ p$_0$ with  $m=+2,+3$.
The frequency $\nu_9$ could be $\ell=3,m=-3$, p$_0$ or $\ell=2, m=0,$ p$_0$ but $\ell=2$ was excluded in our identification for this frequency.
The value of $\nu_7$ is not far from $\ell=2,m=0$, g$_1$ and for $\nu_8$, identified as $\ell=2$, the closest counterpart is $\ell=3, m=+1,$ g$_2$. The frequency $\nu_{10}$, identified as $\ell=1$ or 2, is very close to $\ell=1, m=+1,$ p$_1$.

In Table\,\ref{tab_f}, we compare the theoretical values of the $f$-parameter for MODEL-$6\nu$ with the empirical counterparts
derived with the LTE model atmospheres. The first and second lines for each frequency correspond to the empirical and theoretical values, respectively.
There is no $\nu_7$ because it was not found in the radial velocity variations.
For the frequencies $\nu_6$, $\nu_9$ and $\nu_{10}$, we gave the two values of $f$  obtained for mode degrees, $\ell$, as listed in Table\,1.
The differences in $f$ resulting from adopting different model atmospheres are small; the $f$-parameter derived with the non-LTE
atmospheres is usually slightly lower. In the first column of Table\,\ref{tab_f}, we list the empirical values of the intrinsic mode amplitude, $\varepsilon$,
multiplied by the spherical harmonic, $|\tilde\varepsilon|=|\varepsilon Y_\ell^m(i,0)|$, where $i$ is the inclination angle.

For the radial mode, $\nu_4$, the value of $|\tilde\varepsilon|$ is equal to $|\varepsilon|$, i.e., to
the root mean square of the the relative change of the stellar radius, $\delta r/R$, over the surface.
Because we have identification of the mode degree, $\ell$, and azimuthal order, $m$, for $\nu_1$, $\nu_2$, $\nu_3$, $\nu_5$ and $\nu_A$ as well as
the inclination angle, $i\approx 48^{\rm o}$, \citep{desmet2009}, we were able to derive the relative change of the stellar radius also for these modes.
As a result we obtained that $|\varepsilon|$ is equal to:
1.57\% for $\nu_1$, 0.57\% for $\nu_2$, 0.32\% for $\nu_3$, 0.63\% for $\nu_4$, 0.25\% for $\nu_5$ (assuming that $\ell=2,m=2$) and 0.06\% for $\nu_A$.
The amplitude of the relative radius changes caused by radial pulsations of 12 Lac is similar to that of other $\beta$ Cep stars.
For example, $|\varepsilon|$ of the radial mode for $\theta$ Oph is equal to $0.14\%$ \citep{daszynska_walczak2009},
for $\nu$ Eri - $1.72\%$ \citep{daszynska_walczak2010} and for $\gamma$ Peg - 0.26\% \citep{WDD10}.
The last column contains the value of the instability parameter $\eta$. Some of these modes are stable ($\eta<0$).
The most negative value is for the high-order g-mode, $\nu_A$, and amounts to $-0.63$, what asserts the well-known problem of excitation of high-order g-modes
in the massive main sequence pulsators.
\begin{table}
\begin{center}
\caption{The empirical values of $|\tilde\varepsilon|=|\varepsilon| Y_\ell^m(i,0)$, the empirical and theoretical values of the $f$-parameter,
the instability, $\eta$, and the value of $\chi_E^2$ for all detected frequencies of 12 Lac. We used the stellar model fitting six frequencies:
$\nu_1$, $\nu_2$, $\nu_3$, $\nu_4$, $\nu_5$ and $\nu_A$ (MODEL-$6\nu$). The empirical values of $f$ were derived with the LTE model atmospheres.}
\begin{tabular}{c@{} c@{} c@{} c@{} c@{} c@{} c@{} c@{} c@{}}
\hline
mode &   &$|\tilde\varepsilon|$  & $f_R$  &  $f_I$& $\eta$ \\
\hline
\multirow{2}{1.2cm}{$\nu_{1}$,$\ell=1	$} &	empir~ & 0.01431(62)&-8.41(37)&-1.24(38)& --- \\
&teor&------&-7.97&$-0.11$ &~0.05	\\\hline

\multirow{2}{1.2cm}{$\nu_{2}$,$\ell=1	$} &	empir & 0.00660(37)&-7.69(44)&-0.20(44)& --- \\
&teor&------&-7.77&$-0.23$&~0.053	\\\hline

\multirow{2}{1.2cm}{$\nu_{3}$,$\ell=2	$} &	empir & 0.00429(41)&-10.7(1.3)&-0.6(1.3)& --- \\
&teor&------&-8.62&0.23&0.041	\\\hline

\multirow{2}{1.2cm}{$\nu_{4}$,$\ell=0	$} &	empir & 0.00631(22)&-8.23(33)&-0.31(33)& --- \\
&teor&------&-8.17&0.06&0.044	\\\hline

\multirow{2}{1.2cm}{$\nu_{5}$,$\ell=2	$} &	empir & 0.00187(14)&-4.53(75)&-4.96(75)& --- \\
&teor&------&-6.52	& $-0.90$ &0.087	\\\hline
\multirow{2}{1.2cm}{$\nu_{6}$,$\ell=1	$} &	empir & 0.00029(6)&-20.6(4.0)&4.4(4.0)& --- \\
&teor&------&-11.59&3.21&-0.148	\\

\multirow{2}{1.2cm}{$\nu_{6}$,$\ell=2	$} &	empir & 0.00046(6)&-23.5(3.1)&4.2(3.1)& --- \\
&teor&------&-11.71	&	3.21&-0.153	\\\hline
\multirow{2}{1.2cm}{$\nu_{8}$,$\ell=2	$} &	empir & 0.00163(30)&5.1(1.9)&4.7(1.9)& --- \\
&teor&------&-8.17	&$-0.07$&0.052	\\\hline
\multirow{2}{1.2cm}{$\nu_{9}$,$\ell=1	$} &	empir & 0.00070(12)&-6.0(1.3)&3.3(1.3)& --- \\
&teor&------&-10.53	&	1.96&-0.056	\\

\multirow{2}{1.2cm}{$\nu_{9}$,$\ell=3	$} &	empir & 0.00300(15)&~-13.3(1.8)&23.2(1.8)& --- \\
&teor&------&-10.78&1.91&-0.059	\\\hline
\multirow{2}{1.2cm}{$\nu_{10}$,$\ell=1	$} &	empir & 0.00056(10)&-7.2(1.3)&-0.9(1.4)& --- \\
&teor&------&-9.11&$0.68$&0.02	\\
\multirow{2}{1.2cm}{$\nu_{10}$,$\ell=2	$} &	empir & 0.00074(17)&-4.4(2.2)&-3.0(2.2)& --- \\
&teor&------&-9.20&0.66&0.023	\\\hline
\multirow{2}{1.2cm}{$\nu_{A}$,$\ell=1	$} &	empir & 0.00053(25) & ~26.3(12.5) & ~-2.9(12.0) & --- \\
&teor&------&18.13&5.37&~-0.631	\\\hline
\end{tabular}
\label{tab_f}
\end{center}
\end{table}

The last issue we would like to examine is the effect of the opacity enhancement. The reason is that for 12 Lac, as well as for some other $\beta$ Cephei stars,
there was postulated an increase of stellar opacities near the $Z$-bump \citep{PHD04,MBMD07,DP08}. This should have solved the problem with instability
of some modes as well as improve fitting frequencies. We decided to test the artificial opacity enhancement near this opacity bump. To this aim,
we increased the opacity coefficient in the OPAL tables by 50\% in the metal bump.
The obtained seismic models fitting two frequencies, $\nu_4$ and $\nu_2$, were not changed much. The masses, effective temperatures and luminosities were similar.
As one could expect, the biggest difference is related to the instability. With higher opacities, all our models have unstable modes $\ell=0$, p$_1$
and $\ell=1$, g$_1$, even for $Z=0.007$.
It turned out, that models calculated with the modified OPAL table mimic those computed with the original OP data. It is connected with the internal
differences between the two opacity tables; the OP data provide larger opacity coefficient near the $Z$-bump.

However, with the modified opacities we did not manage to find models fitting the empirical values of the $f$-parameter, neither for $\nu_4$ nor for $\nu_2$.
The increase of the opacity coefficient changed significantly the real part $f$ whereas the imaginary part of $f$ was almost insensitive
to the opacity increase. Although the increase of the opacity parameter can solve problems with instabilities, a disagreement between
the empirical and theoretical values of $f$ argues against this change. The same result was obtained for $\gamma$ Pegasi
(Walczak \& Daszy\'nska-Daszkiewicz 2010, Walczak et al. 2013).

\section{Conclusions}

The aim of the paper was to construct complex seismic models of the $\beta$ Cep/SPB star 12 Lacertae, i.e., models which fit simultaneously
pulsational frequencies and the corresponding values of the nonadiabatic parameter $f$. As usually, we began with the mode identification
from the multi-colour photometry and radial velocity data.
Unambiguous determination of $\ell$ was possible for the six frequencies of 12 Lac; for the remaining ones usually two possibilities were obtained.
Moreover, for the dominant mode, $\nu_1$, and the high-order g-mode, $\nu_A$, the effects of rotation on identification of $\ell$ were checked.
As a result we got that $\nu_1$ is the pure (non-rotational coupled) dipole mode and $\nu_A$ is the dipole retrograde mode.

We found plenty of models which reproduce the two centroid frequencies $\nu_4$ and $\nu_2$, identified as $\ell=0$, p$_1$ and $\ell=1$, g$_1$, respectively,
and their empirical values of the $f$-parameter simultaneously. These models are inside the observational error box only if the OP opacity tables were used.
Models computet with the OPAL data had too small effective temperatures and luminosities.
The next argument for the OP tables is that with this data,
there exist models which fit additionally the high-order g-mode frequency, $\nu_A$, identified as $\ell=1, m=-1$, g$_{13}$, and its values of $f$ for quite low values
of the core overshooting, $\alpha_{\rm ov}\approx 0.2$. With the OPAL data such models exist only for $\alpha_{\rm ov}\approx 0.5$.
Thus, we managed to find models which reproduce the three frequencies and their $f$-parameters simultaneously. This is the first star for which such a good concordance
was achieved.
Moreover, we showed that if one assumes the rotational velocity of the order of 75 km/s for the rotational splitting of the gravity and mixed modes,
then there exist models fitting the four frequencies and a model which fits the six frequencies (the first five and the low frequency $\nu_A$).
This value of the rotational velocity agrees with the result obtained from identification of $\ell$ and $m$ for the high-order g-mode frequency $\nu_A$.
It confirms also the rotational rate suggested by Dziembowski \& Pamyatnykh (2008) for the interior of 12 Lac.

Similarly to the case of the hybrid pulsator $\gamma$ Pegasi, we got that an increase of opacities in the metal bump spoils an agreement between
the empirical and theoretical values of the non-adiabatic parameter $f$. Thus, our conclusion is the same: if modifications of the opacity tables are needed
they have to be done in a more detailed way.

\section*{Acknowledgments}
We thank Gelard Handler for providing us with data on the light variations in the $uvy$ passbands
and Maarten Desmet for data on the radial velocity variations.
This work was partially supported by the Human Capital Programme grant financed by the European Social Fund.


\begin{thebibliography}{99}
\bibitem[\protect\citeauthoryear{Adams}{1912}]{adams} Adams W.S., 1912,
ApJ, 35, 163
\bibitem[\protect\citeauthoryear{Asplund et al.}{2009}]{asplund2009} Asplund M., Grevesse N., Sauval A.J., Scott P., 2009, ARA\&A, 47, 481
 \bibitem[\protect\citeauthoryear{Baglin et al.}{2006}]{Corot}Baglin A., Michel E., Auvergne M., The COROT Team 2006, in Proc. SOHO 18/GONG 2006/HELAS I, Beyond the spherical Sun, ESA SP, Sheffield, 624
\bibitem[\protect\citeauthoryear{Barning}{1963}]{barning} Barning F. J. M., 1963, Bull. Astr. Inst. Netherlands, 17, 22
\bibitem[\protect\citeauthoryear{Christie}{1926}]{christie} Christie W. H., 1926, Pop. Astr., 34, 551
\bibitem[\protect\citeauthoryear{Ciurla}{1987}]{ciurla} Ciurla T., 1987, AcA, 37, 53
\bibitem[\protect\citeauthoryear{Claret}{2000}] {claret2000} Claret A., 2000, A\&A, 363, 1081
\bibitem[\protect\citeauthoryear{Daszy\'nska-Daszkiewicz et al.}{2002}]{ddetal2002} Daszy\'nska-Daszkiewicz J., Dziembowski W. A., Pamyatnykh A. A., Goupil, M.-J.,
2002, A\&A, 392, 151
\bibitem[\protect\citeauthoryear{Daszy\'nska-Daszkiewicz, Dziembowski \& Pamyatnykh}{2003}]{ddp2003} Daszy\'nska-Daszkiewicz J., Dziembowski W. A., Pamyatnykh A. A., 2003, A\&A, 407, 999
\bibitem[\protect\citeauthoryear{Daszy\'nska-Daszkiewicz, Dziembowski \& Pamyatnykh}{2005}]{ddp2005} Daszy\'nska-Daszkiewicz J., Dziembowski W. A., Pamyatnykh A. A., 2005, A\&A, 441, 641
\bibitem[\protect\citeauthoryear{Daszy\'nska-Daszkiewicz \& Szewczuk}{2011}]{dsz2011} Daszy\'nska-Daszkiewicz J., Szewczuk W., 2011, ApJ, 728, 17
\bibitem[\protect\citeauthoryear{Daszy\'nska-Daszkiewicz \& Walczak}{2009}]{daszynska_walczak2009} Daszy\'nska-Daszkiewicz J., Walczak P., 2009, MNRAS, 398, 1961
\bibitem[\protect\citeauthoryear{Daszy\'nska-Daszkiewicz \& Walczak}{2010}]{daszynska_walczak2010} Daszy\'nska-Daszkiewicz J., Walczak P., 2010, MNRAS, 403, 496
\bibitem[\protect\citeauthoryear{de Jager}{1953}]{dejager1953} de Jager C., 1953, BAN, 12, 81
\bibitem[\protect\citeauthoryear{de Jager}{1963}]{dejager} de Jager C., 1963, Bull. Astr. Inst. Netherlands, 17, 1
\bibitem[\protect\citeauthoryear{Desmet et al.}{2009}]{desmet2009} Desmet M., Briquet M., Thoul A., Zima W., De Cat P., Handler G., Ilyin I., Kambe E., Krzesinski J., Lehmann H., Masuda S., Mathias P., Mkrtichian D. E., Telting J., Uytterhoeven K., Yang S.L.S., Aerts C., 2009, MNRAS, 396, 1460
\bibitem[\protect\citeauthoryear{Dziembowski}{1977}]{dziembowski1977} Dziembowski W. A., 1977, AcA, 27, 203
\bibitem[\protect\citeauthoryear{Dziembowski \& Jerzykiewicz}{1999}]{dziembowski_jerzykiewicz1999} Dziembowski W. A., Jerzykiewicz M., 1999, A\&A, 341, 480
\bibitem[\protect\citeauthoryear{Dziembowski \& Pamyatnykh}{2008}]{DP08} Dziembowski W. A., Pamyatnykh A. A., 2008, MNRAS, 385, 2061
\bibitem[\protect\citeauthoryear{Fath}{1938}]{fath1938} Fath E. A., 1938, Pop. Astr., 46, 241
\bibitem[\protect\citeauthoryear{Fath}{1947}]{fath1947} Fath E. A., 1947, AJ, 52, 123
\bibitem[\protect\citeauthoryear{Ferguson at al.}{2005}]{Fal05} Ferguson, J. W., Alexander, D.R., Allard, F., 2005, ApJ, 623, 585
\bibitem[\protect\citeauthoryear{Grevesse \& Noels}{1993}]{grevesse} Grevesse N., Noels A., 1993, in Pratzo N., Vangioni-Flam E., Casse M., eds, Origin and Evolution of the Elements. Cambridge
Univ. Press, Cambridge, p. 15
\bibitem[\protect\citeauthoryear{Guthnick}{1919}]{guthnick} Guthnick P., 1919, AN, 208,219
\bibitem[\protect\citeauthoryear{Handler et al.}{2006}]{handler2006} Handler, G., Jerzykiewicz, M., Rodr\'iguez, E., at al., 2006 MNRAS, 365, 327
\bibitem[\protect\citeauthoryear{Iglesias \&  Rogers}{1996}]{opal} Iglesias C. A., Rogers F. J., 1996, ApJ, 464, 943
\bibitem[\protect\citeauthoryear{Jarzebowski}{1979}]{jarzebowski} Jarzebowski T., Jerzykiewicz M., Musielok B., Le Contel J.-M., 1979, AcA, 29, 517
\bibitem[\protect\citeauthoryear{Jerzykiewicz}{1978}]{jerzykiewicz1978} Jerzykiewicz M., 1978, AcA, 28, 465
\bibitem[\protect\citeauthoryear{Jerzykiewicz}{1984}]{jerzykiewicz1984} Jerzykiewicz M., Musielok B., Borkowski K. J., 1984, AcA, 34, 21
\bibitem[\protect\citeauthoryear{Koch et al.}{2010}]{Kepler} Koch D. G., Borucki  W. J., Basri G.: 2010, ApJ 713, 79
\bibitem[\protect\citeauthoryear{Kurucz}{2004}]{kurucz} Kurucz R. L., 2004, http:// kurucz.harvard.edu
\bibitem[\protect\citeauthoryear{Lanz \& Hubeny}{2007}]{hubeny} Lanz T., Hubeny I., 2007, ApJS, 169, 83
\bibitem[\protect\citeauthoryear{Lee \& Saio}{1997}]{leesaio} Lee, U. \& Saio, H., 1997, ApJ., 491, 839
\bibitem[\protect\citeauthoryear{Mathias et al.}{1994}]{mathias1994} Mathias P., Aerts C., Gillet D., Waelkens C., 1994, A\&A, 289, 875
\bibitem[\protect\citeauthoryear{Miglio et. al}{2007}]{MBMD07} Miglio, A., Bourge, P.-O., Montalb\'an, J., Dupret, M.-A., 2007, CoAst, 150, 209
\bibitem[\protect\citeauthoryear{Morel et al.}{2006}]{morel2006} Morel T., Butler K., Aerts C., Neiner C., Briquet M., 2006, A\&A, 457, 651
\bibitem[\protect\citeauthoryear{Pamyatnykh et al.}{1998}]{pamyatnykh1998} Pamyatnykh A. A., Dziembowski W. A., Handler G., Pikall H., 1998, A\&A, 333, 141
\bibitem[\protect\citeauthoryear{Pamyatnykh, Handler \& Dziembowski}{2004}]{PHD04}Pamyatnykh A. A., Handler G., Dziembowski W. A., 2004, MNRAS, 350, 1022
\bibitem[\protect\citeauthoryear{Rogers \& Nayfonov}{2002}]{RN02} Rogers F. J., Nayfonov A., 2002, ApJ, 576, 1064
\bibitem[\protect\citeauthoryear{Sato}{1979}]{sato} Sato N., 1979, Ap\&SS, 66, 309
\bibitem[\protect\citeauthoryear{Seaton}{2005}]{op} Seaton M. J., 2005, MNRAS, 362, 1
\bibitem[\protect\citeauthoryear{Serenelli at al.}{2009}]{Sal09} Serenelli A. M.. Basu S., Ferguson J. W., Asplund, M., 2009, ApJ, 705, 123
\bibitem[\protect\citeauthoryear{Smith}{1980}]{smith} Smith M. A., 1980, ApJ, 240, 149
\bibitem[\protect\citeauthoryear{Soufi}{1998}]{soufi98} Soufi, F., Goupil, M.-J., Dziembowski, W. A., 1998, A\&A, 334, 911
\bibitem[\protect\citeauthoryear{Stebbins}{1917}]{stebbins} Stebbins J., 1917, Pop. Astr., 25, 657
\bibitem[\protect\citeauthoryear{Townsend}{2003}]{townsend} Townsend, R.H.D. 2003, MNRAS, 340, 1020.
\bibitem[\protect\citeauthoryear{Young}{1915}]{young} Young R. K., 1915, Publ. Dom. Obs. Ottawa, 3, 63
\bibitem[\protect\citeauthoryear{Young}{1919}]{young1919} Young R. K., 1919, JRASC, 13, 45
\bibitem[\protect\citeauthoryear{Walczak \& Daszy\'nska-Daszkiewicz}{2010}]{WDD10} Walczak P. \& Daszy\'nska-Daszkiewicz J., 2010, AN, 331, 1057
\bibitem[\protect\citeauthoryear{Walczak et al.}{2013}]{Wetal13} Walczak P., Daszy\'nska-Daszkiewicz J., Pamyatnykh A. A., Zdravkov T., 2013, in preparation
\bibitem[\protect\citeauthoryear{Walker et al.}{2003}]{MOST} Walker G., Matthews J., Kuschnig R. at al.: 2003, PASP 115, 1023

\end{thebibliography}
\end{document}